\documentclass[journal]{IEEEtran}

\usepackage{latexsym}
\usepackage{graphicx}
\usepackage{amsfonts,amssymb,amsmath}
\usepackage{ctable} 
\usepackage{mathtools, cuted}
\usepackage{hyperref}
\usepackage[ruled,linesnumbered,noend]{algorithm2e}
\usepackage[T1]{fontenc}
\usepackage{cite}
\usepackage{subcaption}
\usepackage{comment}
\usepackage{diagbox}

\usepackage{amsthm}
\usepackage{overpic}
\usepackage{steinmetz}
\usepackage{array}
\usepackage{url}
\usepackage{color}

\usepackage{algorithmicx}
\usepackage{algcompatible}

\usepackage{xcolor}
\usepackage{soul}

\usepackage{flushend}
\usepackage{linegoal}

\theoremstyle{plain}

\newtheorem{lemma}{Lemma}
\newtheorem{example}{Example}

\newcommand{\argmin}[1]{{\underset{{#1}}{\mathrm{arg\,min}}}}
\newcommand{\vect}[1]{\mathbf{#1}}
\newcommand{\maximize}[1]{{\underset{{#1}}{\mathrm{maximize}}}}
\newcommand{\minimize}[1]{{\underset{{#1}}{\mathrm{minimize}}}}

\newcommand*{\LongState}[1]{\STATE
\parbox[t]{0.9\linewidth-\algorithmicindent-\algorithmicindent}{#1\strut}}

\def\diag{\mathrm{diag}}

\def\Htran{\mbox{\tiny $\mathrm{H}$}}
\def\Ttran{\mbox{\tiny $\mathrm{T}$}}
\def\CN{\mathcal{N}_{\mathbb{C}}} 

\SetCommentSty{mycommfont}

\begin{document}

\title{Joint Discrete Precoding and RIS Optimization for RIS-Assisted MU-MIMO Communication Systems}

\author{\IEEEauthorblockN{\normalsize Parisa Ramezani, \textit{Member, IEEE},
    Yasaman Khorsandmanesh, \textit{Student Member, IEEE}, Emil Björnson, \textit{Fellow, IEEE}, }
\thanks{
The authors are with KTH Royal Institute of Technology, Stockholm, Sweden (email:\{parram,yasamank,emilbjo\}@kth.se). This work was supported by the FFL18-0277 grant from the Swedish Foundation for Strategic Research and by the Knut and Alice Wallenberg Foundation. }}

\maketitle
\begin{abstract}
This paper considers a multi-user multiple-input multiple-output (MU-MIMO) system where the downlink communication between a base station (BS) and multiple user equipments (UEs) is aided by a reconfigurable intelligent surface (RIS). We study the sum rate maximization problem with the objective of finding the optimal precoding vectors and RIS configuration. Due to fronthaul limitation, each entry of the precoding vectors must be picked from a finite set of quantization labels. Furthermore, two scenarios for the RIS are investigated, one with continuous infinite-resolution reflection coefficients and another with discrete finite-resolution reflection coefficients. A novel framework is developed which, in contrast to the common literature that only offers sub-optimal solutions for optimization of discrete variables, is able to find the optimal solution to problems involving discrete constraints. 
 Based on the classical weighted minimum mean square error (WMMSE), we transform the original problem into an equivalent weighted sum mean square error (MSE) minimization problem and solve it iteratively. 
We compute the optimal precoding vectors via an efficient algorithm inspired by sphere decoding (SD). 
For optimizing the discrete RIS configuration, two solutions based on the SD algorithm are developed: An optimal SD-based algorithm and a low-complexity heuristic method that can efficiently obtain RIS configuration without much loss in optimality. The effectiveness of the presented algorithms is corroborated via numerical simulations where it is shown that the proposed designs are remarkably superior to the commonly used benchmarks. 
\end{abstract}
\section{Introduction}

To address the rapidly increasing demand for data traffic, it is essential to develop new technologies that can deal with the growing number of mobile devices and emerging data-intensive applications. Reconfigurable intelligent surface (RIS) is one of the  newly-emerged technological solutions that can enable the manipulation of propagation environment to create favorable propagation conditions \cite{Direnzo2020Smart,Wu2020a,Tang2021}. An RIS is essentially a planar arrangement of numerous reflecting elements, often referred to as meta-atoms, positioned with sub-wavelength spacing between them. Each of these elements can independently modify the impinging waves in order to achieve a collective desired functionality such as signal focusing, interference suppression, channel rank improvement, etc. The art of optimizing the phase shift pattern across the RIS is pivotal, as it enables the shaping of the reflected wavefront in a desired manner. This paper particularly considers how an RIS can enable spatial multiplexing of user devices.

Since its emergence, RIS has been extensively studied by the research community and proved effective for enhancing the performance of wireless networks in various ways \cite{Wu2019,Guo2020,Pan2020multicell}. However, theoretical studies cannot be accredited unless they are conducted in accordance with practical system models. Therefore, despite the considerable research on RIS-assisted communication, the application of the technology to real-life scenarios is still far-fetched and more research based on realistic and practical models is called for. For example, the design variables must be optimized under practical hardware constraints.

A typical 5G base station (BS) is composed of an advanced antenna system (AAS) and a baseband unit (BBU) \cite{asplund2020advanced}. AAS is a component consisting of antennas and their corresponding radio units (RUs) and BBU is responsible for processing received uplink and transmitted downlink data. A digital fronthaul connects the AAS and BBU. The fronthaul connection consists of fiber cables with a limited capacity and needs to transfer a number of digital samples per second that grows with the bandwidth, antennas, and users.
Due to the capacity limitation of the fronthaul, each sample must be compressed to be represented by a limited number of bits. Hence, the entries of the precoding vectors computed at the BBU cannot take any value and must belong to a specified discrete set determined by the fronthaul interface. This makes quantization errors inevitable.
Moreover, the literature has reported the design of RIS with either continuous or discrete phase shifts where the former structure is mainly fabricated using varactor diodes and the latter is designed using positive intrinsic negative (PIN) diodes. Specifically, the design of reflective surfaces with continuous phase shifts 
can be found in \cite{Hum2007} where the RIS reflection coefficients can be continuously controlled by changing the voltage applied on the diode. Discrete reflection coefficient implementations with 1-, 2-, and 3-bit resolutions have been presented in \cite{Kamoda2011}, \cite{dai2020reconfigurable}, and \cite{Rains2023}, respectively, where changing the reflection coefficient is achieved by turning on/off the PIN diodes. Both continuous and discrete RIS reflection coeffients are practically implementable. However, due to hardware complexity of the continuous phase shift design, using discrete phase shifts at the RIS is more practically appealing, especially when an RIS with a very large number of elements is to be built. Hence, RIS is expected to be of finite resolution in practice, where each of its elements can only select phase shifts from a discrete set  and the reflection coefficient of each element accordingly belongs to a discrete set as well. 

In this paper, we study a MU-MIMO downlink communication assisted by an RIS, where the direct links between single-antenna  UEs and the multi-antenna BS are obstructed.  The downlink data is calculated at the BBU and is sent to the AAS through the limited-capacity fronthaul infrastructure. Our objective is to design joint optimal precoding matrix and RIS configuration for sum rate maximization. 
 This paper aims to find practical downlink precoding and RIS configurations that adhere to the aforementioned hardware constraints.

\subsection{Related Work}

The literature has extensively addressed the impact of impairments in analog hardware on the communication performance of multi-user multiple-input multiple-output (MU-MIMO) systems, as evidenced by studies such as \cite{Zhang2012a, Bjornson2014a,  aghdam2020distortion}. Additionally, there is a line of work focusing on quantization distortion arising from low-resolution analog-to-digital converters (ADC) in the uplink \cite{mollen2016uplink} and low-resolution digital-to-analog converters (DAC) in the downlink \cite{mezghani2009transmit, jacobsson2019linear}. A notable aspect of these previous works is that the distortion occurs in the RU, i.e., in the analog domain or in the converters. This implies that the transmit signal, obtained after precoding, undergoes distortion. This differs from the setup considered in this paper, where the precoding vectors are subject to distortion because they are computed at the BBU and then sent over the fronthaul to the AAS. The influence of limited fronthaul capacity is examined in \cite{parida2018downlink} with focus on finding analytical achievable rate expressions, while the precoding was not quantized. Taking into account the fronthaul limitation, references \cite{khorsandmanesh2023optimized} and \cite{khorsandmanesh2023fronthaul} proposed novel precoding designs that aimed at minimizing the sum mean square error (MSE) and maximizing the sum rate in a MU-MIMO setting, respectively. 

There have been some recent investigations of the impact of discrete RIS configurations \cite{Alexandropoulos2020,Di2020,Wu2020b,Peng2021reconfigurable,HZhang2022}.  The discrete RIS configuration optimization problem is a combinatorial problem and is inherently hard to solve since the number of configurations grows exponentially with the number of RIS elements. Consequently, the existing works on discrete RIS configuration design typically offer sub-optimal solutions. Specifically, one of the most common approaches taken for obtaining the discrete RIS reflection coefficients is to first find the optimal continuous reflection coefficients and then quantize each of them to the nearest value in the set of available coefficients\cite{Alexandropoulos2020,Peng2021reconfigurable,HZhang2022}. Another approach is to alternatively optimize one reflection coefficient keeping the other ones fixed \cite{Wu2020b,Di2020}. Recently in \cite{Ren2023a}, a geometry-based method has been utilized to find the optimal discrete RIS reflection coefficients considering a single-user single-input single-output scenario. The proposed method has linear complexity with respect to the number of RIS elements, but cannot be applied to more complicated cases with multiple users and/or multiple antennas at either side of the communication. Our previous work in \cite{Ramezani2023Novel} was a primer on the optimal design of discrete RIS reflection coefficients where we used the sphere decoding (SD) algorithm, which has been devised for solving mixed-integer least-squares problems\cite{Agrell2002,Hassibi2005,Vikalo2005}, to optimize the RIS reflection coefficients in an uplink RIS-assisted MU-MIMO setup.   

Additionally, there have been some prior works on sum rate maximization in RIS-assisted MU-MIMO systems. Reference \cite{Guo2020} investigated the weighted sum rate maximization problem by designing the BS precoding and RIS configuration. They first used the closed-form fractional programming approach to obtain a more tractable problem and then utilized the block coordinate descent algorithm to solve the resultant problem. The weighted sum rate maximization problem in a multi-cell MIMO communication system was studied in \cite{Pan2020multicell}, where the authors have utilized the weighted minimum mean square error (WMMSE) approach for alternatively optimizing the BS precoding and RIS configuration.
In \cite{Li2020ICCC}, an aerial ground communication scenario was considered where an RIS is deployed to assist the communication between a UAV and multiple ground users. The sum rate maximization problem was studied under this setup aiming to optimize the UAV trajectory and RIS configuration. The authors have employed a conjugate gradient algorithm to optimize the RIS configuration and utilized a successive convex approximation approach to find the UAV trajectory. The authors in \cite{Zhang2023Sum} studied a sum rate maximization problem in a multi-user RIS-assisted system by leveraging the statistical CSI at the BS and used the random matrix theory to derive the asymptotic sum rate. They then provided a new formulation of the sum rate maximization problem with the asymptotic sum rate being the objective function. After decoupling the precoding and RIS optimization problems, a projected gradient ascent algorithm and a water-filling approach were applied to find sub-optimal RIS configuration and precoding vectors, respectively. All the aforementioned works on sum rate maximization in RIS-assisted systems only considered continuous RIS phase shifts. In \cite{Di2020}, Di \emph{et al.} considered discrete phase shifts at the RIS and formulated a mixed-integer sum rate maximization problem. They then proposed an iterative approach for designing the BS precoding and RIS configuration. However, as mentioned earlier, this work only provides a sub-optimal solution for RIS configuration. Besides, none of the above works take the fronthaul limitation at the BS side into account.

\subsection{Contributions}
We study the joint precoding and RIS configuration design in practical RIS-assisted MU-MIMO communication systems. We start by assuming a continuous infinite-resolution RIS and then extend our design to the practical discrete RIS. We utilize an SD-based algorithm to design the optimal precoding vectors and discrete RIS configuration. This proposed framework for optimizing discrete precoding and RIS configurations is a major step toward operating the technology efficiently under practical hardware limitations. The main contributions of this paper are as outlined below:  
\begin{itemize}
    \item We formulate a sum rate maximization problem for optimizing the BS precoding vectors and RIS configuration. Two RIS configuration scenarios are considered, one with continuous infinite-resolution reflection coefficients and another with discrete finite-resolution reflection coefficients, where in the latter scenario, only a few reflection coefficients are available for RIS elements to choose from.  
     We also take into account the fronthaul capacity limitation when solving the sum rate maximization problems, necessitating the precoding entries to be selected from a discrete set. The discrete constraints on the RIS configuration and precoding make the problem a non-convex combinatorial optimization problem.  Though joint precoding and RIS configuration design has been studied in some prior works (e.g., \cite{Gan2021}), this is the first work that considers both discrete precoding and finite-resolution RIS configurations.
     We transform the original sum rate maximization problem into an equivalent weighted sum MSE minimization problem and adopt an alternating optimization approach based on the block coordinate descent method to solve it. 
    \item We propose a solution based on the Schnorr Euchner SD (SESD) algorithm to solve the precoding optimization sub-problem. This solution provides the global optimal solution to the precoding optimization sub-problem while significantly reducing the complexity compared to an exhaustive search. We further employ the SESD algorithm to solve the RIS configuration optimization problem when RIS reflection coefficients belong to a discrete set. In contrast to the existing literature, where only sub-optimal solutions for discrete RIS configuration are offered, the SESD algorithm finds the optimal discrete RIS configurations. This is of paramount importance in a MU-MIMO RIS-assisted system, where sub-optimal configuration of the RIS limits the interference suppression capability of the system and severely degrades the performance.
    \item Considering the large dimension of the RIS and the exponential complexity of the standard SESD algorithm with respect to the number of RIS elements, we devise a novel low-complexity heuristic algorithm that sequentially optimizes the reflection coefficients of subsets of RIS elements using the SESD algorithm. By properly choosing the number of RIS elements in each subset, one can achieve the desired trade-off between complexity and optimality. To the authors’ best knowledge, this is the first work to present an efficient heuristic algorithm based on SESD that, by reducing the problem dimension, can solve mixed-integer least-squares problems with sufficiently low complexity to allow arrays with many elements.
    \item We evaluate all the proposed algorithms via numerical simulations under different system setups and show the superiority of the presented SESD-based designs over the commonly used benchmarks. We also assess the effectiveness of the proposed low-complexity heuristic SESD algorithm for RIS configuration optimization and show that it performs close to the optimal SESD design.
\end{itemize}

\subsection{Organization and Notations}
\textit{Organization:} We describe the system model and formulate the sum rate maximization problem in Section~\ref{sec:system_model}. The reformulation of the sum rate maximization problem as a weighted sum MSE minimization problem and the solution to this problem are presented in Section~\ref{sec:proposed_WMMSE} considering continuous infinite-resolution configuration at the RIS. The scenario of discrete finite-resolution configuration at the RIS is investigated in Section~\ref{sec:RIS_discrete} where two solutions based on the SD methodology are developed for optimizing the RIS configuration. We analyze the complexity of our proposed solutions in Section~\ref{sec:complexity}. Numerical results are provided in Section~\ref{sec:Num_results} and Section~\ref{sec:conclusions} concludes the paper. 

\textit{Notations:} Scalars are denoted by italic letters, vectors and matrices are denoted by bold-face lower-case and upper-case letters, respectively. For a complex-valued vector $\vect{x}$, $\|\vect{x}\|$ denotes its norm, $[\vect{x}]_m$ is its $m$th entry, $\vect{x}[m:n]$ is the vector containing entries $m$ to $n$ of $\vect{x}$, and $\diag(\vect{x})$ represents a diagonal matrix having $\vect{x}$ on its main diagonal. Furthermore, $[\vect{X}]_{m,n}$ shows the element in $m$th row and $n$th column of the matrix $\vect{X}$, $\vect{X}[m_1:n_1,m_2:n_2]$ indicates the matrix formed by taking the elements in rows $m_1$ to $n_1$ and columns $m_2$ to $n_2$ of $\vect{X}$, and $\diag(\vect{X})$ returns the vector which is formed by taking the elements on the main diagonal of $\vect{X}$. $\Re(\cdot)$ and $\Im(\cdot)$ denote the real and imaginary parts of a complex number, vector, or matrix. $\mathbb{E}\{X\}$ represents the statistical average of a random variable $X$. $(\cdot)^*$, $(\cdot)^{\Ttran}$, and $(\cdot)^{\Htran}$ indicate the conjugate, transpose, and conjugate transpose, respectively.  
$\mathbb{R}$ and $\mathbb{C}$ denote the set of real and complex numbers, respectively, and 
$\CN (0,\sigma^2)$ indicates a circularly symmetric complex Gaussian distribution with variance $\sigma^2$.

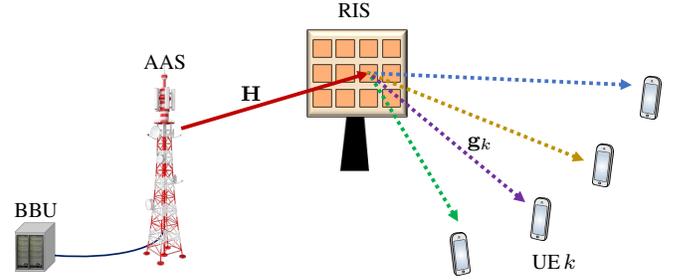
\begin{figure}[t!]
  \centering
   \begin{overpic}[scale = 0.28]{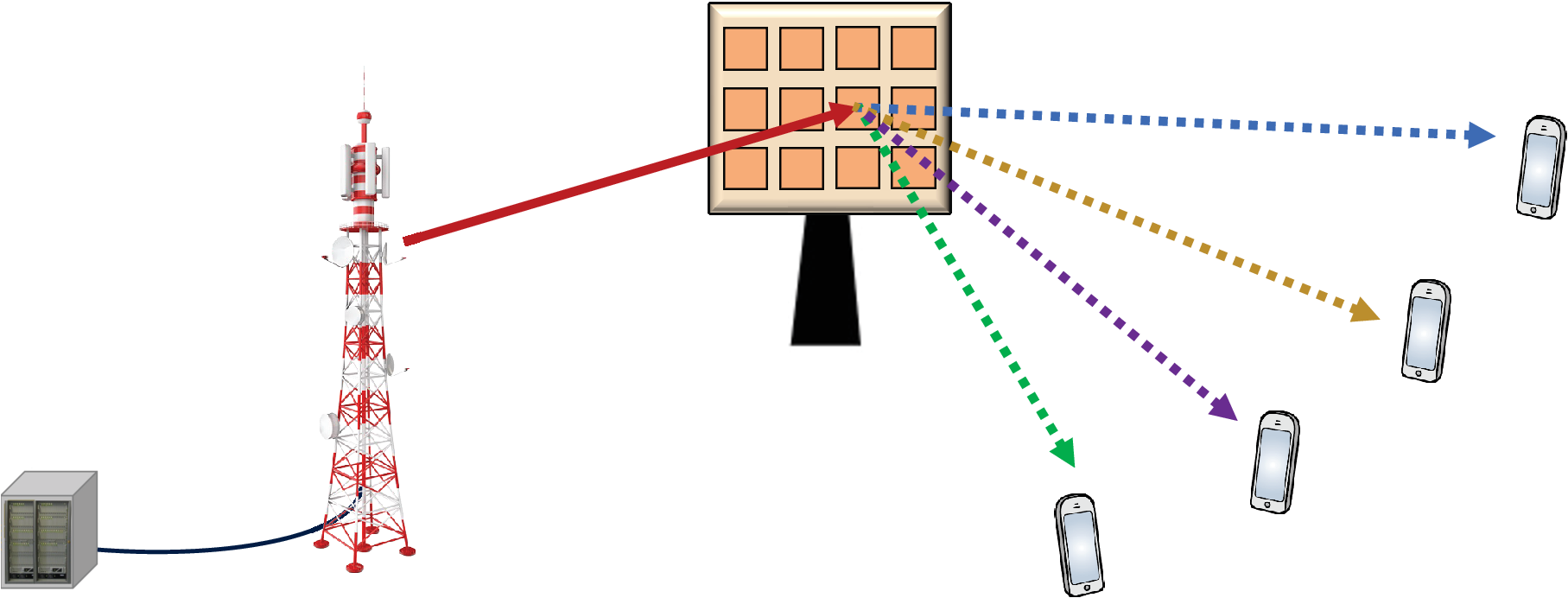}
  \put(0,9){\footnotesize BBU}%
  \put(20,32){\footnotesize AAS}%
  \put(50,40){\footnotesize RIS}%
  \put(80,1){\footnotesize UE\,$k$}%
  \put(35,27){\footnotesize $\vect{H}$}
  \put(70,20){\footnotesize $\vect{g}_k$}
   \end{overpic}
\caption{An RIS-assisted MU-MIMO downlink communication.}
\label{fig:system}
\end{figure}

\section{System Model and Problem Formulation}
\label{sec:system_model}
We consider an RIS-assisted downlink communication system, as depicted in Fig.~\ref{fig:system}, where a BS equipped with an AAS with $M$ antenna-integrated radios communicates with $K$ single-antenna user equipments (UEs) through an RIS with $N$ reflecting elements. The direct links between the BS and the UEs are assumed to be blocked or severely degraded and the BS can only serve the UEs via the RIS. The AAS is connected to a BBU through a fronthaul link with limited capacity, modeled as a finite-resolution quantizer. The signal transmitted by the BS to each UE consists of a precoding vector and a data symbol.
Specifically, the signal intended for UE\,$k$ is $\vect{w}_k s_k$, where $\vect{w}_k \in \mathbb{C}^M$ represents the linear precoding vector for UE\,$k$ and $s_k$ is the unit-power data symbol. The BBU encodes the data and computes the precoding matrix and sends the data and precoding vector separately to the AAS through the limited-capacity fronthaul link. Since the data symbols are bit sequences picked from a channel coding codebook, the BBU can send them to the AAS via the fronthaul link without quantization errors.
However, the precoding vectors contain arbitrary complex-valued entries and must be quantized due to the digital fronthaul. After receiving the data symbols and the precoding vectors, the AAS maps data symbols to modulation symbols, multiplies them with their corresponding precoding vectors and transmits $\sum_{k=1}^K \vect{w}_k s_k$ in the downlink. The transmitted signal must satisfy the following power constraint 
\begin{equation}
\label{eq:sum_power_constraint}
   \mathbb{E}\left\{\|\vect{w}_k s_k\|^2\right\} = \sum_{k=1}^K \|\vect{w}_k\|^2 \leq p,
\end{equation}
where $p$ is the maximum average transmit power of downlink signals and the equality comes from the fact that the transmitted symbols are independent and identically distributed and have unit power.

The received signal at UE\,$k$ can be expressed as 
\begin{equation}
\label{eq:received_signal}
  y_k = \vect{g}_k^{\Ttran}\boldsymbol{\Theta} \vect{H}\vect{w}_k s_k + \sum_{i=1,i\neq k}^K \vect{g}_k^{\Ttran}\boldsymbol{\Theta} \vect{H}\vect{w}_i s_i + n_k,  
\end{equation}
where $\vect{g}_k \in \mathbb{C}^{N}$ and $\vect{H} \in \mathbb{C}^{N\times M}$ denote the channel from the RIS to UE\,$k$ and from the BS to the RIS, respectively. $\boldsymbol{\Theta} = \diag (\theta_1,\theta_2,\ldots,\theta_N)$ is the RIS reflection coefficient matrix with 
$\theta_n = e^{j\vartheta_n}$ and $\vartheta_n$ being the phase shift induced by the $n$th element to the impinging signal. Finally, $n_k \sim \CN (0,N_0)$ indicates the independent additive
complex Gaussian receiver noise with power $N_0$. We assume the availability of perfect channel state information (CSI) in this paper to characterize the maximum performance that can be achieved in the RIS-assisted MU-MIMO communication system using the proposed techniques. CSI acquisition can be done using the techniques proposed in \cite{He2020Cascaded,Zhou2021Joint,Haghshenas2024Parametric,Swindlehurst2022Channel}.

The achievable rate of UE\,$k$ is obtained as $\log_2\big(1+\mathrm{SINR}_k\big)$, where $\mathrm{SINR}_k$ is the signal-to-interference-plus-noise ratio (SINR) at UE\,$k$ which, in our considered setup, is given by 
\begin{equation}
    \mathrm{SINR}_k (\vect{W},\boldsymbol{\theta}) = \frac{|\boldsymbol{\theta}^{\Ttran} \vect{F}_k\vect{w}_k|^2}{\sum_{i=1,i\neq k}^K |\boldsymbol{\theta}^{\Ttran}\vect{F}_k \vect{w}_i|^2 + N_0},
\end{equation}with $\vect{W} = [\vect{w}_1,\ldots,\vect{w}_K]$, $\vect{F}_k = \diag(\vect{g}_k)\vect{H}$, and $\boldsymbol{\theta} = \diag(\boldsymbol{\Theta})$.

\subsection{Fronthaul Quantization}
\label{sec:fronthaul_quantization}
The entries of the precoding vector $\vect{w}_k,~k=1,\ldots,K,$ belong to a finite alphabet set that is defined as 
\begin{equation}
\label{eq:quantization_alphabet}
    \mathcal{P} = \{l_R + jl_I: l_R,l_I \in \mathcal{L}\}.
\end{equation}
We assume that the same quantization alphabet is used for both real and imaginary components. Note that $\mathcal{P}$ is equal to the set of complex numbers $\mathbb{C}$ in case of infinite resolution. As uniform quantization is normally used in practice, we model our quantizer $\mathcal{Q(\cdot)}: \mathbb{C} \rightarrow \mathcal{P}$ as a symmetric uniform quantization with step size $\Delta$. The quantizer function $\mathcal{Q(\cdot)}$ can be uniquely defined by the set of quantization labels $\mathcal{L} = \{l_0,\ldots,l_{L-1}\}$ and the set of quantization thresholds $\mathcal{T} = \{\tau_0,\ldots,\tau_L\}$ where $L = |\mathcal{L}|$ is the number of quantization levels. In particular, each entry of $\mathcal{L}$ is defined as 
\begin{equation}
   l_i = \Delta \left(i - \frac{L-1}{2}\right),~~i = 0,\ldots, L-1. 
\end{equation}
Furthermore, we have $-\infty =\tau_0 < \tau_1 <\ldots <\tau_{L-1} < \tau_L = \infty$ as quantization thresholds where 
\begin{equation}
    \tau_i = \Delta \left( i - \frac{L}{2}\right),~~i = 1,\ldots,L-1.
\end{equation}
A complex input $x = x_{\mathrm{R}} + jx_{\mathrm{I}}$ is mapped to $\mathcal{Q}(x) = l_x + jl_y$ if $x_{\mathrm{R}} \in [\tau_x, \tau_{x+1})$ and $x_{\mathrm{I}} \in [\tau_y,\tau_{y+1})$. The step size of the quantizer $\Delta$ should
be chosen to minimize the distortion between the quantized output and the unquantized input.  The optimal $\Delta$ depends on the statistical distribution of the input, which in our case depends
on the precoding scheme and the channel model. Since the distribution of the precoding matrix
elements is generally unknown and varies with the user population, we set the step size to
minimize the distortion under the maximum-entropy assumption that the per-antenna input to
the quantizers is distributed as $\CN(0,\frac{p}{KM})$ where the variance is selected so that the sum power
of the elements matches with the power constraint in \eqref{eq:sum_power_constraint}. The corresponding optimal step size
for the normal distribution was found in \cite{Hui2001}.

\subsection{Problem Formulation}
 We are interested in maximizing the sum rate of the considered downlink communication system under the maximum transmit power constraint in \eqref{eq:sum_power_constraint} at the BS and unit modulus constraints on the RIS reflection coefficients. We formulate the problem as 
\begin{subequations}
\begin{align}
\label{eq:main_problem}
~~ &\maximize{\vect{W} \in \mathcal{P}^{M \times K},\boldsymbol{\theta} \in \mathbb{C}^{N} }\quad &&\sum_{k=1}^K\log_2\,\big(1+\mathrm{SINR}_k(\vect{W},\boldsymbol{\theta})\big) \\
 &\quad\mathrm{subject~to}\,\quad &&\sum_{k=1}^K \|\vect{w}_k\|^2 \leq p, \label{eq:power_constraint}\\
 &~&& |\theta_n| = 1,~n=1,\ldots,N. \label{eq:unit_modulus}
 \end{align}
 \end{subequations}
 Problem (7) is highly non-convex due to the fractional structure of the SINR expression, non-concave objective function. coupled variables, and the unit modulus constraints in \eqref{eq:unit_modulus}. This problem is very challenging to solve in its original form. 
 Furthermore, taking into account the fronthaul limitation for precoding design makes the problem even more challenging to solve and as will be discussed in Section~\ref{sec:RIS_discrete}, considering discrete RIS reflection coefficients will further add to the intractability of the problem. Therefore, finding the global optimal solution is difficult. In this paper, we target a locally optimal solution by re-formulating problem (7) into a more tractable form. 

 \section{Iterative Weighted Sum MSE Minimization}
 \label{sec:proposed_WMMSE}
 Inspired by the classical WMMSE approach, we utilize the well-known equivalence between sum rate maximization and weighted sum MSE minimization problems \cite{Shi2011} which allows us to split the main problem into sub-problems and alternately solve them in an iterative manner. In this section, we assume continuous RIS configurations; we will consider the more practical discrete RIS configuration scenario in the next section.

 Assume that $\hat{s}_k = \beta_k y_k$ is the estimate of the symbol $s_k$ at UE\,$k$ where $\beta_k$ denotes the receiver gain (also known as precoding factor \cite{Jacobsson2017} and beamforming factor \cite{POKEMON}), and is used at UE\,$k$ to estimate the transmitted symbol. For a given receiver gain, the MSE at UE\,$k$ is computed as 
 \begin{align}
         \label{eq:MSE}
         e_k &= \mathbb{E}\{|s_k - \hat{s}_k|^2\} \notag \\ &=|\beta_k|^2 \left( |\boldsymbol{\theta}^{\Ttran}\vect{F}_k\vect{w}_k|^2 + \sum_{i=1,i\neq k}^K |\boldsymbol{\theta}^{\Ttran}\vect{F}_k \vect{w}_i|^2 + N_0\right) \notag \\ & - 2\Re\left(\beta_k \boldsymbol{\theta}^{\Ttran}\vect{F}_k \vect{w}_k\right) + 1 .
 \end{align}
We can see that when either the precoding vector $\vect{w}_k$ or the RIS configuration $\boldsymbol{\theta}$ is held fixed, the MSE expression is a convex function with respect to the other variable. This will help us re-write problem (7) in a more tractable form.

The following lemma 
presents a reformulation of the sum rate maximization problem (7) based on weighted sum MSE minimization which is adapted from \cite{Shi2011} to the problem at hand. 

\begin{lemma}
\label{lem:WMMSE}
   Let $c_k$ be an auxiliary weight associated with UE\,$k$. The sum rate maximization problem (7) is equivalent to the following problem 
   \begin{align}
 \label{eq:WMMSE} &\minimize{\substack{ \vect{W} \in \mathcal{P}^{M\times K}\\ \boldsymbol{\theta} \in \mathbb{C}^N, \boldsymbol{\beta} \in \mathbb{C}^K, \vect{c} \in \mathbb{R}^K}}\,\,\, \left(\sum_{k=1}^K c_k e_k - \log_2 (c_k) \right),\\
    &\quad \mathrm{subject~to}\,\,  \eqref{eq:power_constraint} \notag, \eqref{eq:unit_modulus},    
    \end{align}  in the sense that the optimal precoding $\vect{W}$ and optimal RIS configuration vector $\boldsymbol{\theta}$ are the same for both problems. 
   In problem \eqref{eq:WMMSE}, $\vect{c} = [c_1,\ldots,c_K]^{\Ttran}$ and $\boldsymbol{\beta} = [\beta_1,\ldots,\beta_K]^{\Ttran}$.
   \end{lemma}

   \begin{IEEEproof}
       Given the BS precoding vectors and RIS configuration, the optimal receiver gain $\overline{\beta}_k$ can be obtained by minimizing the MSE expression in \eqref{eq:MSE}, which is a quadratic function of $\beta_k$. The optimal solution is obtained as 
   \begin{equation}
   \label{eq:receiver_gain}
       \overline{\beta}_k = \frac{\left(\boldsymbol{\theta}^{\Ttran}\vect{F}_k \vect{w}_k\right)^*}{|\boldsymbol{\theta}^{\Ttran}\vect{F}_k \vect{w}_k|^2 + \sum_{i=1,i\neq k}^K |\boldsymbol{\theta}^{\Ttran}\vect{F}_k \vect{w}_i|^2 + N_0}.
   \end{equation}
 Furthermore, the objective function of \eqref{eq:WMMSE} is convex with respect to $c_k$ and its optimal value can be obtained by taking the derivative as 
 
 \begin{equation}
 \label{eq:weights}
    \overline{c}_k = \frac{1}{\ln (2)e_k} = \frac{1}{\ln(2)} \left( 1 + \frac{|\boldsymbol{\theta}^{\Ttran}\vect{F}_k \vect{w}_k|^2}{\sum_{i=1,i\neq k} |\boldsymbol{\theta}^{\Ttran}\vect{F}_k \vect{w}_i|^2 + N_0}\right). 
 \end{equation}
 If we substitute \eqref{eq:weights} into the objective function of \eqref{eq:WMMSE}, the problem for optimizing the precoding $\vect{W}$ and RIS configuration $\boldsymbol{\theta}$ will be given by 
 \begin{equation}
    \label{eq:optimize_w_theta}
  \minimize{\substack{ \vect{W} \in \mathcal{P}^{M\times K}\\ \boldsymbol{\theta} \in \mathbb{C}^N}} - \sum_{k=1}^K  \log_2 \left(\frac{1}{e_k} \right) = \maximize{\substack{ \vect{W} \in \mathcal{P}^{M\times K}\\ \boldsymbol{\theta} \in \mathbb{C}^N}} \sum_{k=1}^K  \log_2 \left(\frac{1}{e_k} \right),
 \end{equation}subject to \eqref{eq:power_constraint} and \eqref{eq:unit_modulus}.  
 After substituting the optimal receiver gains in \eqref{eq:receiver_gain} into the MSE expression in \eqref{eq:MSE}, we arrive at 
\begin{align}
  \frac{1}{e_k} &= \frac{\sum_{i=1}^K \left|\boldsymbol{\theta}^{\Ttran} \vect{F}_k \vect{w}_i \right|^2 +N_0}{\sum_{i=1,i\neq k}^K \left|\boldsymbol{\theta}^{\Ttran} \vect{F}_k \vect{w}_i \right|^2 +N_0} \nonumber \\
  & = 1 + \frac{\left|\boldsymbol{\theta}^{\Ttran}\vect{F}_k \vect{w}_k\right|^2}{\sum_{i=1,i\neq k}^K \left|\boldsymbol{\theta}^{\Ttran} \vect{F}_k \vect{w}_i \right|^2 +N_0} = 1+ \mathrm{SINR}_k(\vect{W},\boldsymbol{\theta}).
\end{align} Therefore, \eqref{eq:optimize_w_theta} is equivalent to \eqref{eq:main_problem} and problem \eqref{eq:WMMSE} is equivalent to problem (7).
   \end{IEEEproof}

This lemma shows that the sum rate maximization problem in (7) is equivalent to the weighted sum MSE minimization problem in \eqref{eq:WMMSE}, in the sense that the optimal precoding vectors and RIS configuration are the same for the two problems. While the sum rate expression in  \eqref{eq:main_problem} is non-concave with respect to precoding vectors and RIS configuration, the objective function of problem~\eqref{eq:WMMSE} is a convex function of all the optimization variables. This motivates the use of a block coordinate descent approach where the variables are divided into separate blocks and alternately optimized in an iterative manner.

 \subsection{Optimizing Precoding Vectors}
 \label{sec:optimize_precoding}
 Fixing the receiver gains, associated weights, and RIS phase configuration, the problem of finding the optimal BS precoding vectors is given by 
     \begin{align}
         \label{eq:precoding_optimization} & \minimize{ \vect{W} \in \mathcal{P}^{M\times K}}\,\,\, \sum_{k=1}^K c_k e_k,\\
    &\mathrm{subject~to}\,\, \eqref{eq:power_constraint}. \notag
     \end{align}
 Dropping the constant terms from the MSE expression, the objective function in \eqref{eq:precoding_optimization} can be expressed as 
 \begin{equation}
    \sum_{k=1}^K c_k |\beta_k|^2 |\vect{f}_k^{\Ttran} \vect{w}_k|^2 + \sum_{i=1,i\neq k}^K c_i |\beta_i|^2 |\vect{f}_i^{\Ttran}\vect{w}_k|^2 - 2c_k\Re\left(\beta_k \vect{f}_k^{\Ttran}\vect{w}_k\right), 
 \end{equation}
 where $\vect{f}_k^{\Ttran} = \boldsymbol{\theta}^{\Ttran}\vect{F}_k$. We take a Lagrange multiplier approach to reformulate problem \eqref{eq:precoding_optimization}. In particular, using the Lagrange multiplier $\mu$, the Lagrangian function can be written as 
    \begin{align}
    \label{eq:Lagrangian}
      \mathcal{F}(\vect{W},\mu) &= \sum_{k=1}^K  \vect{w}_k^{\Htran} \Big( \sum_{i=1}^K c_i |\beta_i|^2 \vect{f}_i^* \vect{f}_i^{\Ttran} + \mu \vect{I}_M \Big)\vect{w}_k \notag \\ & - (c_k\beta_k\vect{f}_k^{\Ttran})\vect{w}_k - \vect{w}_k^{\Htran}(c_k \beta_k \vect{f}_k^{\Ttran})^{\Htran} - \mu p.
    \end{align}
For a fixed value of $\mu$, the problem for minimizing the Lagrangian in \eqref{eq:Lagrangian} with respect to $\vect{W}$ can be split into $K$ independent sub-problems, each solved for one of the precoding vectors. Defining $\vect{V} = \sum_{i=1}^K c_i |\beta_i|^2 \vect{f}_i^* \vect{f}_i^{\Ttran} + \mu \vect{I}_M$ and $\vect{v}_k = c_k \beta_k \vect{f}_k$, the sub-problem for optimizing $\vect{w}_k$ can be expressed as 
\begin{equation}
\label{eq:optimize_wk}
    \minimize{\vect{w}_k \in \mathcal{P}^M}\,\,\,\, \vect{w}_k^{\Htran} \vect{V} \vect{w}_k - \vect{v}_k^{\Ttran}\vect{w}_k - (\vect{v}_k^{\Ttran}\vect{w}_k)^{\Htran},
\end{equation}
If we neglect the fronthaul limitation and assume infinite-resolution precoding, problem \eqref{eq:optimize_wk} would become a convex problem for which the optimal solution is readily found by setting the derivative of the quadratic objective function with respect to $\vect{w}_k$ to zero. Specifically, the infinite-resolution precoding vectors are obtained as 

\begin{equation}
\label{eq:infinite_precoding}
    \vect{w}_k^{\mathrm{inf}} = \vect{V}^{-1} \vect{v}_k^*.
\end{equation}
Therefore, a simple approach to find the discrete precoding vectors is to first optimize $\vect{w}_k$ by dropping the fronthaul constraint and then use the quantizer $\mathcal{Q}(\cdot)$ (defined in Section~\ref{sec:fronthaul_quantization}) to map the real and imaginary parts of each entry of the optimal continuous precoding vector to their corresponding nearest label in $\mathcal{L}$. This is a heuristic sub-optimal solution to \eqref{eq:optimize_wk} which leads to extra interference and reduced beamforming gains
due to separate quantization of the entries of the precoding vectors. An optimal design of the precoding can reduce the interference by tuning the quantization errors so that they cancel each other.

Instead of following the above two-step procedure where the continuous ideal
precoding vectors are first computed and then quantized, we herein present an efficient method based on the SD algorithm to directly compute the per-UE precoding vectors whose elements come from the given discrete alphabet. SD algorithm provides the globally optimum or maximum likelihood solution of the integer least-squares problem with a much lower complexity than exhaustive search \cite{Viterbo1999,Agrell2002,Han2011ICC}.
Specifically, we propose to use SESD algorithm \cite{Agrell2002}, which improves upon the basic SD algorithm by initially examining the smallest child node of each parent node in each layer, as the problem is a depth-first tree search. Here, child nodes are all possible nodes in the precoding vector starting from its last entry, and then their parent's nodes are the upper entries of the possible precoding vector in a hierarchical way. This strategy exploits the fact that the first feasible solution found is often highly suitable, enabling a rapid reduction in the search radius. Consequently, numerous branches can be pruned, resulting in a further reduction of computational complexity. To apply SESD, we rewrite the objective function in \eqref{eq:optimize_wk} as

\begin{algorithm}[t]
\caption{SESD algorithm for solving \eqref{eq:SD_real}.}
\label{Alg:SESD}
\begin{algorithmic}[1]
\STATEx {\textbf{Inputs}:  $\Tilde{\vect{R}}_w$, $\tilde{\vect{d}}_{w,k}$, $\mathcal{L}$, $2M$. }
\STATE{$\tilde{\vect{w}}_k \leftarrow \vect{0}_{2M \times 1}$, $\check{\vect{w}}_k \leftarrow \vect{0}_{2M \times 1}$, $m \leftarrow 2M+1$, $\textbf{r} \leftarrow \vect{0}_{2M \times 1}$, $ r_{\mathrm{opt}} \leftarrow \mathrm{inf}$, $\mathrm{state} \leftarrow  \mathrm{down}$, $\mathrm{flag} \leftarrow 0$}
\WHILE{$!\Big( (\mathrm{state} == \mathrm{up}) \& \& (m == 2M) \Big)$}
\IF{$(\mathrm{state} == \mathrm{down})$}
\STATE{$m \leftarrow m-1$}
\LongState{Compute \begin{small}$\vect{z}_m[l] =  
  \big|\Tilde{\vect{d}}_{w,k}[m] - \Tilde{\vect{R}}_w[m,m+1:2M]\check{\vect{w}}_k[m+1:2M] - $   $\Tilde{\vect{R}}_w[m,m]\mathcal{L}(l)\big|,\,\forall l = 1,\ldots,L$\end{small}} 
\LongState{Sort $\vect{z}_m$ in ascending order and collect the sorting indices in the vector $\vect{p}_m$, i.e., $[\sim,\vect{p_m}] = \mathrm{sort}(\vect{z}_m$)}
\STATE{$s_m \leftarrow 1$} 
\LongState{$\check{\vect{w}}_k[m] \leftarrow \mathcal{L}\left(\vect{p}_m (s_m)\right)$ \tcp*[f]{ Select the label corresponding  to the minimum value in $\vect{z}_m$}}   
\STATE{$s_m = s_m +1$}
\ELSE
\STATE{$m \leftarrow m+1$}
\IF{$s_m > L$}
\LongState{$\mathrm{flag} = 1$ \tcp*[f]{ All labels in this level already checked}}
\ELSE
\LongState{$\mathrm{flag} = 0$,\,$\check{\vect{w}}_k[m] \leftarrow \mathcal{L}\left(\vect{p}_m (s_m)\right)$,\,$s_m = s_m+1$}
\ENDIF
\ENDIF
\IF{$\mathrm{flag} ==1$ }
\STATE{$\mathrm{state} \leftarrow \mathrm{up}$}
\ELSE
\LongState{$\vect{r}[m] \leftarrow \big|\Tilde{\vect{d}}_{w,k} [m] - \tilde{\vect{R}}_w[m,m:2M]\check{\vect{w}}[m:2M]\big|^2 + \vect{r}[m+1] $\tcp*[f]{Compute the partial Euclidean distance}} 
\IF{$\vect{r}[m] < r_{\mathrm{opt}}$ }
\IF{$(m = = 1)$}
\STATE{$\Tilde{\vect{w}}_k\leftarrow\check{\vect{w}}_k$ }
\STATE{$r_{\mathrm{opt}} \leftarrow \vect{r}[m],\, \mathrm{state} \leftarrow \mathrm{up}$}
\ELSE
\STATE{$\mathrm{state} \leftarrow \mathrm{down}$ }
\ENDIF
\ELSE
\STATE{$\mathrm{state} \leftarrow \mathrm{up}$}
\ENDIF
\ENDIF
\ENDWHILE
\STATEx{ \textbf{Output:} $\overline{\vect{w}}_k = \Tilde{\vect{w}}_k [1:M] + j \Tilde{\vect{w}}_k [M+1:2M]$} 
 \end{algorithmic}
\end{algorithm}

\begin{equation}
   \|\vect{d}_{w,k} - \vect{R}_w\vect{w}_k\|^2 - \vect{d}_{w,k}^{\Htran}\vect{d}_{w,k},
\end{equation}
where $\vect{R}_w\in \mathbb{C}^M$ is an upper triangular matrix obtained from the Cholesky decomposition of $\vect{V}$ as $\vect{V} = \vect{R}_w^{\Htran}\vect{R}_w$ and $\vect{d}_{w,k} = (\vect{v}_k^{\Ttran}\vect{R}_w^{-1})^{\Htran}$. Using this notation, problem \eqref{eq:optimize_wk} can be recast as 
\begin{equation}
\label{eq:SD_problem}
    \minimize{\vect{w}_k \in \mathcal{P}^M}\,\,\, \|\vect{d}_{w,k} - \vect{R}_w\vect{w}_k\|^2.
\end{equation}
Problem \eqref{eq:SD_problem} is classified as an integer least-squares problem due to the finite-resolution precoder. In problem \eqref{eq:SD_problem}, the entries of the precoding vectors are complex and the search space contains complex-valued alphabets. Noting that the real and imaginary parts of the set $\mathcal{P}$ are independent, we can write problem \eqref{eq:SD_problem} in an equivalent real-valued form. This way, the run time of the SESD algorithm will be reduced because the two-dimensional search for the complex precoding entries will turn into a linear search. Therefore, we define 
\begin{equation}
\begin{aligned}
    &\tilde{\vect{d}}_{w,k} = \begin{bmatrix}
    \Re(\vect{d}_{w,k}) \\
     \Im(\vect{d}_{w,k}),
    \end{bmatrix},~ \tilde{\vect{w}}_k = \begin{bmatrix}
    \Re(\vect{w}_k) \\
     \Im(\vect{w}_k) 
    \end{bmatrix}, \\
   & \tilde{\vect{R}}_w =
  \begin{bmatrix}
    \Re(\vect{R}_w) & -\Im(\vect{R}_w)  \\
    \Im(\vect{R}_w)  & \Re(\vect{R}_w)
  \end{bmatrix}.
\end{aligned}
\end{equation}
We can now present the real-valued reformulation of \eqref{eq:SD_problem} as 
\begin{equation}
\label{eq:SD_real}
    \minimize{\tilde{\vect{w}}_k \in \mathcal{L}^{2M}}\,\,\, \|\tilde{\vect{d}}_{w,k} - \tilde{\vect{R}}_w\tilde{\vect{w}}_k\|^2.
\end{equation}
We apply the classical SESD algorithm for solving \eqref{eq:SD_real} for a fixed value of $\mu$. We can find the optimal value of $\mu$ which provides a solution that satisfies \eqref{eq:power_constraint} near equality via a bisection search. We denote the optimal precoding solution by $\overline{\vect{w}}_k$. The pseudo-code of the SESD algorithm for solving \eqref{eq:SD_real} and finding $\overline{\vect{w}}_k$ is provided in Algorithm~\ref{Alg:SESD}. Here is a brief description of the algorithm:
\begin{figure*}[t]
    \centering
    \includegraphics[width = \textwidth]{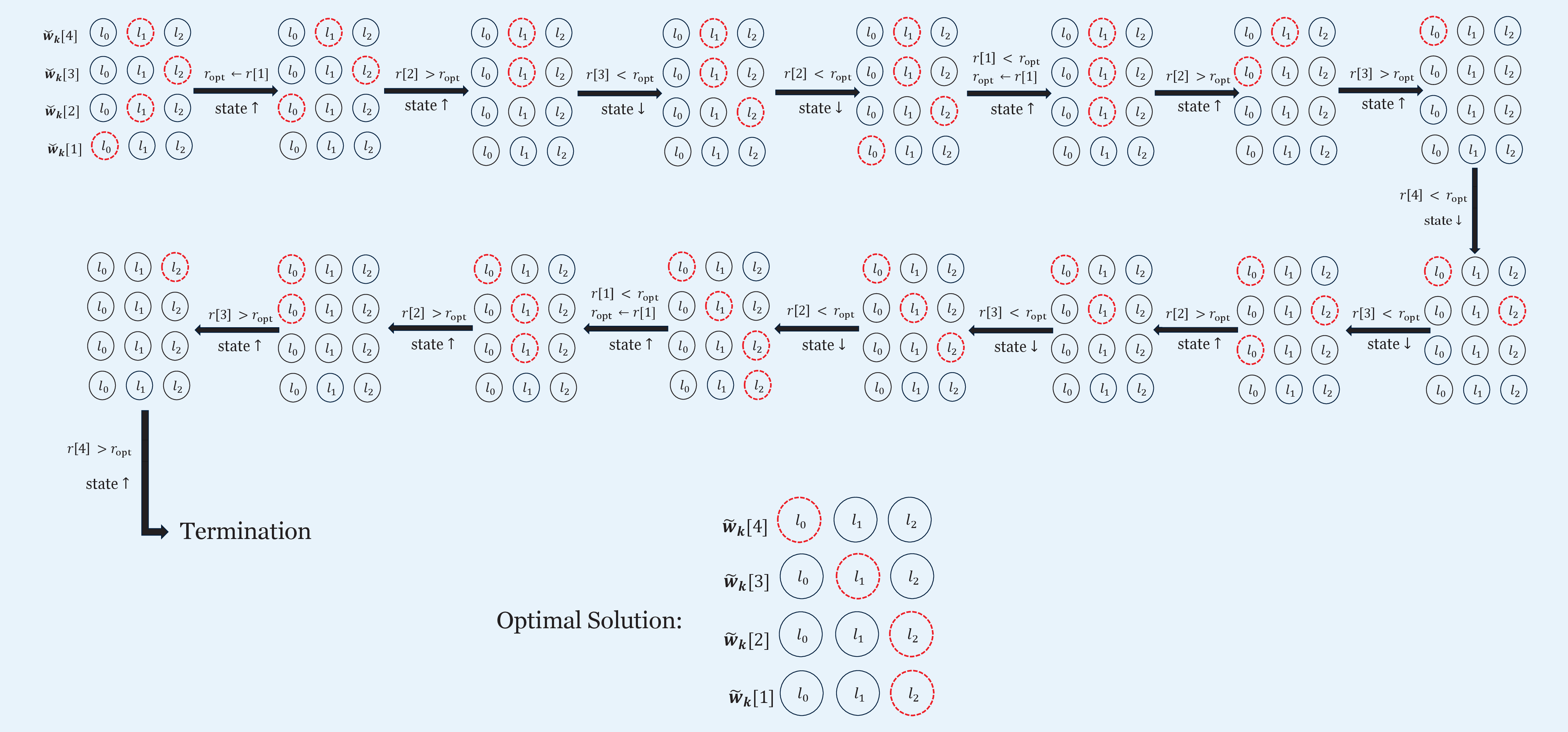}
    \caption{An illustrative example of SESD.}
    \label{fig:SD_example}
\end{figure*}
The initial search radius is set to infinity. Starting from level $m = 2M$, the algorithm first computes $  \left|\Tilde{\vect{d}}_{w,k} [2M] - \tilde{\vect{R}}_w[2M,2M]\vect{w}_k[2M]\right|$ for all $\vect{w}_k[2M] \in \mathcal{L}$. It sorts the computed values, stores the sorting indices in $\vect{p}_{2M}$ and selects the quantization label which minimizes the above absolute value as $\check{\vect{w}}_k[2M]$. In other words, at this stage, $\check{\vect{w}}_k[2M] = \mathcal{L}(\vect{p}_{2M}[1])$. Then, it proceeds to the lower level and finds $\vect{p}_{2M-1}$ and $\check{\vect{w}}_k[2M-1]$ in the same way given the value of $\check{\vect{w}}_k[2M]$. It then continues the process for all other levels.  Specifically, at level $m$, the sorting indices are stored in $\vect{p}_m$ and $\check{\vect{w}}_k[m] = \mathcal{L}(\vect{p}_m[1])$ is chosen. The chosen value of $\check{\vect{w}}_k[m]$ essentially equals to 

\begin{align}
\label{eq:precoding_minimizer}
 &\check{\vect{w}}_k[m] =  \argmin{\vect{w}_k[m] \in \mathcal{L}}\left| \xi_m -  \Tilde{\vect{R}}_w[m,m]\vect{w}_k[m]\right|,  
\end{align}where
\begin{equation}
    \xi_m = \Tilde{\vect{d}}_{w,k}[m] - \Tilde{\vect{R}}_w[m,m+1:2M]\check{\vect{w}}_k[m+1:2M].
\end{equation}

After solving \eqref{eq:precoding_minimizer} for all $2M$ levels, the obtained combination of precoding labels, i.e., $\check{\vect{w}}_k$, is selected as the temporary optimal solution and stored in $\Tilde{\vect{w}}_k$. Furthermore, the radius given by 
\begin{equation}
   \vect{r}[1] =  \sum_{m=1}^{2M} \big| \Tilde{\vect{d}}_{w,k}[m] - \Tilde{\vect{R}}_w[m,m:2M]\check{\vect{w}}_k[m:2M]\big|^2
\end{equation}
is chosen as the new search radius. Then, fixing $\check{\vect{w}}_k[3], \ldots, \check{\vect{w}}_k[2M]$, the algorithm  updates $\check{\vect{w}}_k[2]$ by selecting another label, i.e., $\mathcal{L}(\vect{p}_2[2])$ and checks if \begin{small}$\vect{r}[2] = \sum_{i=2}^{2M} \left| \tilde{\vect{d}}_{w,k} [i] - \tilde{\vect{R}}_k[i,i:2M]\Tilde{\vect{w}}_k[i:2M]   \right|^2$\end{small} with the newly updated $\check{\vect{w}}_k[2]$ exceeds the search radius $r_{\mathrm{opt}}$ or not. The reason why the algorithm does not update $\check{\vect{w}}_k[1]$ and directly goes for updating $\check{\vect{w}}_k[2]$ is that for the combination of entries $\check{\vect{w}}_k[2],\ldots, \check{\vect{w}}_k[2M]$, the value $\check{\vect{w}}_k[1]$ obtained by solving \eqref{eq:precoding_minimizer} for $m = 1$ is the optimal choice among all the quantization labels and results in the smallest possible radius; thus, there is no need to explore other labels in the set at this level. If $\vect{r}[2] < r_{\mathrm{opt}}$, the algorithm proceeds to update $\check{\vect{w}}_k[1]$ by solving \eqref{eq:precoding_minimizer} again.
Otherwise, it moves to an upper level and updates $\check{\vect{w}}_k[3]$ having $\check{\vect{w}}_k[4],\ldots,\check{\vect{w}}_k[2M]$ fixed. This procedure is repeated and after each successful update of $\check{\vect{w}}_k[m]$ (i.e., the update that results in $\vect{r}[m]<r_{\mathrm{opt}}$), the algorithm moves to a lower level and finds $\check{\vect{w}}_k[m-1]$ from \eqref{eq:precoding_minimizer}. When updating $\check{\vect{w}}_k[m]$ is unsuccessful, the value of $m$ is increased by 1 and $\check{\vect{w}}_k[m+1]$ is updated by choosing another label from $\mathcal{L}$ based on the sorting indices at this level. Whenever a new combination of entries $\check{\vect{w}}_k[1],\ldots,\check{\vect{w}}_k[2M]$ is found that satisfies  $\vect{r}[1] = \|\Tilde{\vect{d}}_{w,k} - \Tilde{\vect{R}}_w \check{\vect{w}}_k\|^2 < r_{\mathrm{opt}}$, the search radius is updated as $r_{\mathrm{opt}} = \vect{r}[1]$. It is important to note the role of sorting indices in updating the precoding entries. Specifically, assuming that $\check{\vect{w}}_k[m] = \mathcal{L}(\vect{p}_m[i])$, the next time the algorithm moves from level $m-1$ to level $m$, $\check{\vect{w}}_k[m] = \mathcal{L}(\vect{p}_m[i+1])$ is selected as the new precoding entry at this level. If  $\check{\vect{w}}_k[m] = \mathcal{L}(\vect{p}_m[i+1])$ fails to fall in the optimal radius $r_{\mathrm{opt}}$, there is no need for further update of $\check{\vect{w}}_k[m]$ for the given values of $\check{\vect{w}}_k[m+1],\ldots,\check{\vect{w}}_k[2M]$. In other words, if $\vect{r}[m] > r_{\mathrm{opt}} $ for $\check{\vect{w}}_k[m] = \mathcal{L}(\vect{p}_m[i+1])$, we will surely have $\vect{r}[m] > r_{\mathrm{opt}} $ for  $\check{\vect{w}}_k[m] = \mathcal{L}(\vect{p}_m[i^\prime]),\,i^\prime > i+1$.

To further clarify how the SESD algorithm works, we present an illustrative example in Fig.~\ref{fig:SD_example}. In this example, $\mathcal{L} = \{l_0,l_1,l_2\}$ and $2M = 4$; thus, the objective is to optimize the entries of $\Tilde{\vect{w}}_k \in \mathbb{C}^{4}$ where for each entry, there are three possible options: $l_0,l_1,l_2$. The starting point in the figure is $\check{\vect{w}}_k = [l_1,l_2,l_1,l_0]^{\Ttran}$ where each entry is assumed to be obtained by solving \eqref{eq:precoding_minimizer}. Dashed red circles show the selected points from $\mathcal{L}$ for each level. The search radius is set as $r_{\mathrm{opt}} = \vect{r}[1]$ and the algorithm goes one level up to look for another feasible $\check{\vect{w}}_k[2]$.  In this example,  $\check{\vect{w}}_k[2]$ is changed from $l_1$ to $l_0$ and it is then checked if the new combination of entries falls within the optimal search radius (i.e., if $\vect{r}[2] < r_{\mathrm{opt}}$). Each time the algorithm finds $\vect{r}[m+1] < r_{\mathrm{opt}}$, it searches for a new $\check{\vect{w}}_k[m]$ given $\check{\vect{w}}_k[m+1],\ldots,\Tilde{\vect{w}}_k[2M]$. If $m = 1$ and the combination of entries falls within the optimal search radius, the optimal radius is updated. Whenever $\vect{r}[m]>r_{\mathrm{opt}}$, the algorithm goes to an upper level and looks for a new $\check{\vect{w}}_k[m+1]$. The figure exemplifies this search process which ends with returning $\Tilde{\vect{w}}_k = [l_0,l_1,l_2,l_2]^{\Ttran}$ as the final optimal solution.   

\begin{example}
Consider the following problem
\begin{equation}
\label{eq:example_problem}
    \minimize{\tilde{\vect{w}}_k \in \mathcal{L}^4}\, \left\| \tilde{\vect{d}}_{w,k} - \tilde{\vect{R}}_w \tilde{\vect{w}}_k\right\|^2,
\end{equation}
where the set of labels is $\mathcal{L} = \{-1,1,2\}$ and 
    \begin{equation}
    \tilde{\vect{d}}_{w,k} = \begin{bmatrix}
    2  \\
    3 \\
    1 \\
    3  
  \end{bmatrix},~
\tilde{\vect{R}}_w =
  \begin{bmatrix}
    16 & 2 & 3 & 13  \\
     0 & 11 & 10 & 8  \\
      0 & 0 & 6 & 12 \\
       0 & 0 & 0 & 1   
  \end{bmatrix}. 
\end{equation} The initial vector  found by Algorithm \ref{Alg:SESD} is $\check{\vect{w}}_k = [-1, -1, -1, 2]^{\Ttran}$ which is obtained by solving \eqref{eq:precoding_minimizer} at each of the four levels. With this vector, we have $\vect{r}[1] = 363$. Then, the search for a vector with a smaller $\vect{r}[1]$ continues. The next vector found by the algorithm is $\check{\vect{w}}_k = [-1,1,-1,1]^{\Ttran}$ with $\vect{r}[1] = 101$. Finally, $\Tilde{\vect{w}}_k = [1,-1,2,-1]^{\Ttran}$ is returned as the optimal solution to \eqref{eq:example_problem} with $\vect{r}[1] = 46$.
\end{example}

\subsection{Optimizing the RIS Configuration}
\label{sec:optimize_RIS}
We now proceed to find the optimal RIS configuration given the receiver gains, associated weights, and precoding vectors. Defining $\vect{u}_{k,i} = \beta_k^* \vect{F}_k^* \vect{w}_i^*$, we rewrite the MSE in \eqref{eq:MSE} as 
\begin{equation}
   e_k =  \boldsymbol{\theta}^{\Htran} \Big( \sum_{i=1}^K \vect{u}_{k,i} \vect{u}_{k,i}^{\Htran}\Big)\boldsymbol{\theta} - \vect{u}_{k,k}^{\Htran} \boldsymbol{\theta} - \boldsymbol{\theta}^{\Htran}\vect{u}_{k,k} + |\beta_k|^2 N_0 + 1.
\end{equation}
Discarding the terms that are independent of $\boldsymbol{\theta}$, the RIS configuration optimization problem can be formulated as 
    \begin{align}
        \label{eq:optimize_theta}
&\minimize{\boldsymbol{\theta} \in \mathbb{C}^N}\,\,\boldsymbol{\theta}^{\Htran} \left( \sum_{k=1}^K c_k \sum_{i=1}^K \vect{u}_{k,i} \vect{u}_{k,i}^{\Htran}\right)\boldsymbol{\theta} \notag \\&~~~~~~~~~~~ - \Big(\sum_{k=1}^K c_k  \vect{u}_{k,k}^{\Htran}\Big) \boldsymbol{\theta} - \boldsymbol{\theta}^{\Htran} \Big(\sum_{k=1}^K c_k \vect{u}_{k,k}\Big) \\
        &\mathrm{subject~to} \,\,\eqref{eq:unit_modulus}. \notag
    \end{align}Despite the quadratic objective function, problem \eqref{eq:optimize_theta} is non-convex due to the unit modulus constraints in \eqref{eq:unit_modulus}. To solve \eqref{eq:optimize_theta}, we adopt an alternating optimization approach and sequentially optimize one of the reflection coefficients while keeping the others fixed. Expanding the objective function of \eqref{eq:optimize_theta}, the problem can be rewritten as 
    \begin{align}
        \label{eq:optimize_theta2}
&\minimize{\boldsymbol{\theta} \in \mathbb{C}^N}\,\,\sum_{\tilde{n} = 1}^N \theta_{\tilde{n}} \sum_{n = 1}^N  \theta_n^* [\vect{A}]_{n,\Tilde{n}} - 2 \Re \left(\sum_{\tilde{n} = 1}^N \theta_{\tilde{n}} [\vect{t}]_{\tilde{n}}^*\right)\\
        &\mathrm{subject~to} \,\,\eqref{eq:unit_modulus}, \notag
    \end{align}where
    \begin{align}
        \label{eq:matrix_A}\vect{A} &= \sum_{k=1}^K c_k \sum_{i=1}^K \vect{u}_{k,i} \vect{u}_{k,i}^{\Htran}, \\
        \vect{t}  &= \sum_{k=1}^K c_k \vect{u}_{k,k}. \label{eq:vector_t}
    \end{align}
Fixing $\theta_n,\, n\neq \hat{n}$, the optimization problem with respect to $\theta_{\hat{n}}$ reduces to 
\begin{equation}
        \label{eq:optimize_one_theta}
        \minimize{\theta_{\hat{n}} = e^{j\vartheta_{\hat{n}}}}\,\,2 \Re \left(\theta_{\hat{n}} \left( \sum_{n \neq \hat{n}} \theta_n^* [\vect{A}]_{n,\hat{n}} - [\vect{t}]^*_{\hat{n}}\right) \right).
\end{equation}
It is straightforward to see that the solution to \eqref{eq:optimize_one_theta} is obtained as 
\begin{equation}
\label{eq:optimal_phase_shift}
    \Bar{\theta}_{\hat{n}} = e^{j\left ( - \arg\left( \sum_{n \neq \hat{n}} \theta_n^* [\vect{A}]_{n,\hat{n}} - [\vect{t}]^*_{\hat{n}}\right) + \pi \right)}.
\end{equation}
We alternately optimize the RIS reflection coefficients using \eqref{eq:optimal_phase_shift} in an iterative fashion until we achieve convergence.

\subsection{Overall Algorithm}
Following the procedure described above, we iteratively optimize the receiver gains, associated weights, BS precoding vectors, and RIS configuration until a satisfactory convergence is attained. Algorithm~\ref{Alg:WMMSE} summarizes the presented iterative weighted sum MSE minimization. For initializing the precoding vectors, we opt for the infinite-resolution regularized zero-forcing precoding which can be derived by minimizing the sum MSE \cite{Joham}, i.e., we set 
\begin{equation}
    \overline{\vect{W}}^{(0)} = \vect{H}^{\Htran} \boldsymbol{\Theta}^{(0)\Htran}\vect{G}^{\Htran} \left( \vect{G}\boldsymbol{\Theta}^{(0)}\vect{H}\left(\vect{G}\boldsymbol{\Theta}^{(0)}\vect{H}\right)^{\Htran} + \frac{KN_0}{p}\vect{I}_K \right)^{-1}, 
\end{equation}where $\vect{G} = [\vect{g}_1,\vect{g}_2,\ldots,\vect{g}_K]^{\Ttran}$. Moreover, $\boldsymbol{\Theta}^{(0)} = \diag(\overline{\boldsymbol{\theta}}^{(0)})$ is the initial RIS configuration matrix. In our simulations, we initialize the RIS configuration randomly.   
\begin{algorithm}[t!]
\caption{Iterative WMMSE algorithm for solving problem \eqref{eq:WMMSE}}
\label{Alg:WMMSE}
\begin{algorithmic}[1]
\STATEx {\textbf{Inputs}: channels $\vect{H}$, $\vect{G}$, noise power $N_0$, number of UEs $K$, maximum average transmit power $p$ }
\STATE{Initialize the RIS configuration $\overline{\boldsymbol{\theta}}^{(0)}$ and BS precoding matrix $\overline{\vect{W}}^{(0)} = [\overline{\vect{w}}_1^{(0)},\ldots,\overline{\vect{w}}_K^{(0)}]$}
\STATE{Compute $\overline{\beta}_k^{(0)} = \frac{\left(\overline{\boldsymbol{\theta}}^{(0)\Ttran}\vect{F}_k \overline{\vect{w}}^{(0)}_k\right)^*}{|\overline{\boldsymbol{\theta}}^{(0)\Ttran}\vect{F}_k \overline{\vect{w}}^{(0)}_k|^2 + \sum_{i=1,i\neq k}^K |\overline{\boldsymbol{\theta}}^{(0)\Ttran}\vect{F}_k \overline{\vect{w}}^{(0)}_i|^2 + N_0}$ }
\STATE{Set $\overline{\vect{c}}^{(0)} = [\overline{c}_1^{(0)},\ldots,\overline{c}_k^{(0)}]^{\Ttran} = \boldsymbol{1}_K$}
\STATE{Set the convergence threshold $\epsilon>0$ and maximum number of iterations $L_{\mathrm{max}}$}
\STATE{Set $\varepsilon = \epsilon + 1$, $l =0$}
\STATE{Given $\overline{\vect{W}}^{(0)}$, $\overline{\boldsymbol{\theta}}^{(0)}$, and $\overline{\boldsymbol{\beta}}^{0}$, find $e_k^{(0)}$ from \eqref{eq:MSE} and compute $f^{(0)} = \sum_{k=1}^K \overline{c}_k^{(0)} e_k^{(0)} - \log_2(\overline{c}_k^{(0)})$}
\WHILE{$\varepsilon > \epsilon$ and $l < L_{\mathrm{max}}$}
\STATE{$l=l+1$}
\LongState{Using Algorithm~\ref{Alg:SESD} and bisection method, find the optimal precoding vectors $\{\overline{\vect{w}}_k^{(l)}\}$}
\LongState{Find  $\overline{\boldsymbol{\theta}}^{(l)}$ by alternating optimization of reflection coefficients using \eqref{eq:optimal_phase_shift}}
\STATE{Obtain the optimal receiver gains $\{\overline{\beta}_k^{(l)}\}$ from \eqref{eq:receiver_gain}}
\STATE{Obtain the optimal weights $\{\overline{c}_k^{(l)}\}$ from \eqref{eq:weights}}
\LongState{Compute $\{e_k^{(l)}\}$ by substituting $\{\overline{\beta}_k^{(l)}\}$, $\{\overline{\vect{w}}_k^{(l)}\}$, and $\overline{\boldsymbol{\theta}}^{(l)}$ into the MSE expression \eqref{eq:MSE}}
\STATE{Compute $f^{(l)} = \sum_{k=1}^K \overline{c}_k^{(l)} e_k^{(l)} - \log_2(\overline{c_k}^{(l)})$}
\STATE{Compute $\varepsilon = |f^{(l)} - f^{(l-1)}|$}
\ENDWHILE
\STATEx{ \textbf{Outputs:} $\overline{\vect{w}}_k^{(l)}~k = 1,\ldots, K,$ and $\overline{\boldsymbol{\theta}}^{(l)}$} 
 \end{algorithmic}
\end{algorithm}

\section{RIS with Finite Phase Shift Resolution}
\label{sec:RIS_discrete}
In practice, due to hardware design limitations, it may not be possible for RIS elements to take any phase value. Therefore, a practical RIS is expected to have a finite phase shift resolution, meaning that only a limited number of phase shifts are available for RIS elements to choose from. In such a case, the discrete set of reflection coefficients is given by 
\begin{equation}
\label{eq:discrete_set}
    \mathcal{D} = \left\{e^{j\frac{m\pi}{2^{b-1}}}: m =0,1,\ldots,2^b - 1 \right\},
\end{equation}
where $b$ is the bit-resolution of the RIS elements. 

We now aim to maximize the downlink sum rate, this time assuming discrete reflection coefficients at the RIS. The optimization problem can thus be expressed as 
    \begin{align}
    \label{eq:main_problem_discrete}
        &\maximize{\vect{W} \in \mathcal{P}^{M \times K},\boldsymbol{\theta} \in \mathcal{D}^N}\quad \,\,\sum_{k=1}^K\log_2\,\big(1+\mathrm{SINR}_k(\vect{W},\boldsymbol{\theta})\big) \\
 &~~~~\mathrm{subject~to}\,\,  \eqref{eq:power_constraint}. \notag
    \end{align}
We take the same iterative steps as before to solve \eqref{eq:main_problem_discrete}. Specifically, we first reformulate \eqref{eq:main_problem_discrete} as an equivalent weighted sum MSE minimization problem similar to the one in \eqref{eq:WMMSE}. We then alternately optimize the blocks of variables until convergence is achieved. In each iteration, the optimal receiver gains and weights are obtained from \eqref{eq:receiver_gain} and \eqref{eq:weights}, respectively, while solving the problem \eqref{eq:SD_real} yields optimal BS precoding vectors. The new aspect is the discrete RIS resolution, which results in a different optimization problem for the RIS configuration:
\begin{equation}
        \label{eq:optimize_theta_discrete}
        \minimize{\boldsymbol{\theta} \in \mathcal{D}^N}\,\,\boldsymbol{\theta}^{\Htran}\vect{A}\boldsymbol{\theta} - \vect{t}^{\Htran}\boldsymbol{\theta} - \boldsymbol{\theta}^{\Htran} \vect{t},
\end{equation}
where $\vect{A}$ and $\vect{t}$ are defined in \eqref{eq:matrix_A} and \eqref{eq:vector_t}, respectively. One way to solve problem \eqref{eq:optimize_theta_discrete}, which is prevalent in the prior literature, is to first solve the problem for the case of a continuous RIS configuration and then map each obtained reflection coefficient to the nearest feasible point in the set $\mathcal{D}$. With this approach, sub-optimal discrete RIS reflection coefficients are attained as 
\begin{equation}
\label{eq:nearest_point}
\tilde{\boldsymbol{\theta}}[n] = \argmin{\boldsymbol{\theta}[n] \in \mathcal{D}}\,\left| \overline{\boldsymbol{\theta}}[n] - \boldsymbol{\theta}[n]\right|,~n = 1,2,\ldots, N,
\end{equation}
where $\overline{\boldsymbol{\theta}}$ is obtained using the method described in Section~\ref{sec:optimize_RIS}. This is a low-complexity approach for finding the discrete RIS configuration; however, it is far from optimal because the reflection coefficients are selected independently and the quantization errors pile up, which is particularly troublesome in multi-user setups where the errors lead to inter-user interference. 

Herein, we propose another method based on the SESD technique to directly find the optimal discrete RIS configuration instead of solving the problem for the continuous configuration case and then quantizing each reflection coefficient. We seek an objective function in the form of the one in problem \eqref{eq:SD_problem}. We thus rewrite the objective function in \eqref{eq:optimize_theta_discrete} as 
\begin{equation}
    \|\vect{d}_{\theta} - \vect{R}_{\theta}\boldsymbol{\theta}\|^2 - \vect{d}_{\theta}^{\Htran}\vect{d}_{\theta}, 
\end{equation}where the upper triangular matrix $\vect{R}_{\theta}$ is obtained from the Cholesky decomposition of $\vect{A}$ as $\vect{A} = \vect{R}_{\theta}^{\Htran}\vect{R}_{\theta}$ and $\vect{d}_{\theta} = (\vect{t}^{\Htran} \vect{R}_{\theta}^{-1})^{\Htran}$. Therefore, problem \eqref{eq:optimize_theta_discrete} is rewritten as 
\begin{equation}
   \label{eq:SD_problem2}\minimize{\boldsymbol{\theta} \in \mathcal{D}^N}\,\, \|\vect{d}_{\theta} - \vect{R}_{\theta}\boldsymbol{\theta}\|^2.  
\end{equation}
The triangular structure of $\vect{R}_{\theta}$ allows us to leverage the SESD algorithm for solving problem \eqref{eq:SD_problem2}. Note that here, we cannot reformulate the SESD problem as an equivalent real-valued form similar to the one in \eqref{eq:SD_real} by separating the real and imaginary parts of $\boldsymbol{\theta}$ because the real and imaginary parts of $\boldsymbol{\theta}$ are intertwined and cannot be separately optimized. Instead of searching over a line as in the precoding optimization problem, we have to search over the unit circle. Therefore, the inputs of Algorithm \ref{Alg:SESD} for optimizing the RIS configuration will be  $\vect{R}_\theta$, $\vect{d}_\theta$, $\mathcal{D}$, and $N$.

\subsection{Rank-Deficiency Problem in RIS Configuration Optimization}
Matrix $\vect{A}$ in \eqref{eq:matrix_A} is a square matrix of size $N$ being formed by the summation of $K^2$ rank-one matrices. Therefore, if $K^2 < N$, then $\vect{A}$ is rank-deficient, the number of RIS reflection coefficients to be optimized is greater than the rank of $\vect{A}$, and the RIS configuration optimization problem is under-determined. In this case, zero-valued elements appear on the diagonal of the upper triangular matrix $\vect{R}_\theta$, making the standard SD algorithms inapplicable.
Note that this issue did not exist in finding the optimal precoding vectors via SESD in Section~\ref{sec:optimize_precoding} because the term $\mu \vect{I}_M$ in the Lagrangian function \eqref{eq:Lagrangian} ensured that matrix $\vect{V}$ has full rank. 

We propose a technique for dealing with the rank-deficiency of $\vect{A}$ which is inspired by the generalized SD algorithm proposed in \cite{cui2004efficient}. In other words, we modify the RIS configuration optimization problem in a way to make it solvable via the standard SESD algorithm. Due to the unit modulus entries of the RIS configuration vector $\boldsymbol{\theta}$, we have $\|\boldsymbol{\theta}\|^2=\boldsymbol{\theta}^{\Htran}\boldsymbol{\theta} = N$. Therefore, adding the constant term $\alpha \|\boldsymbol{\theta}\|^2$ to the objective function of \eqref{eq:optimize_theta_discrete} does not change its optimal solution. Doing so, the problem \eqref{eq:optimize_theta_discrete} is modified as 
\begin{equation}
        \minimize{\boldsymbol{\theta} \in \mathcal{D}^N}\,\,\boldsymbol{\theta}^{\Htran}(\vect{A} + \alpha \vect{I}_N)\boldsymbol{\theta} - \vect{t}^{\Htran}\boldsymbol{\theta} - \boldsymbol{\theta}^{\Htran} \vect{t}.
\end{equation}Consequently, problem \eqref{eq:SD_problem2} turns into 
\begin{equation}
   \label{eq:generalized_SD}\minimize{\boldsymbol{\theta} \in \mathcal{D}^N}\,\, \|\tilde{\vect{d}}_{\theta} - \Tilde{\vect{R}}_{\theta}\boldsymbol{\theta}\|^2, 
\end{equation}where $\vect{A} + \alpha \vect{I}_N = \Tilde{\vect{R}}_{\theta}^{\Htran}\Tilde{\vect{R}}_{\theta}$ and $\Tilde{\vect{d}}_{\theta} = (\vect{t}^{\Htran}\Tilde{\vect{R}}_{\theta}^{-1})^{\Htran}$. Since all the diagonal elements of $\Tilde{\vect{R}}_{\theta}$ are non-zero, the SESD technique can be applied to solve problem \eqref{eq:generalized_SD}.

\vspace{3mm}
\subsection{Low-Complexity Heuristic SESD Algorithm for RIS Configuration Optimization}
\label{sec:heuristic_SESD}

Although SESD drastically reduces the complexity compared to an exhaustive search over all $2^{bN}$ RIS configurations, it still incurs exponential complexity with respect to the RIS size. Specifically, the complexity of solving problem \eqref{eq:SD_problem2} or \eqref{eq:generalized_SD} using the SESD technique is $O(2^{\gamma b N})$ for some $0 \leq \gamma \leq 1$ \cite{jalden2005complexity}. This restricts the usage of SESD for optimizing the discrete configuration of an RIS with a large number of elements. To overcome this limitation, we herein propose a heuristic SESD-based algorithm for discrete RIS configuration optimization which sequentially optimizes small subsets of the RIS reflection coefficients using the SESD algorithm. We can also apply this heuristic SESD-based algorithm to reduce the complexity of solving the precoding optimization problem in Section~\ref{sec:optimize_precoding} if the number of BS antennas is large. 

Consider the problem in \eqref{eq:generalized_SD} where the objective is to optimize the discrete configuration of an RIS with $N$ elements. We split this problem into $\eta$ sub-problems, each aiming to optimize the discrete configuration of $N_\eta = N/\eta$ elements. We start with the last $N_\eta$ elements of the RIS and solve the following problem via the SESD algorithm:
\begin{equation}
\label{eq:heuristic_SD}\minimize{\boldsymbol{\theta} \in \mathcal{D}^{N_\eta}}\,\, \big\|\hat{\vect{d}}_{\theta} - \hat{\vect{R}}_{\theta}\boldsymbol{\theta}\big\|^2, 
\end{equation}where
\begin{align}
   &\hat{\vect{d}}_\theta = \tilde{\vect{d}}_\theta \left[ (\eta -1)N_\eta +1:N \right], \\
   & \hat{\vect{R}}_\theta = \Tilde{\vect{R}}_\theta \left[(\eta -1)N_\eta +1:N, (\eta -1)N_\eta +1:N \right].
\end{align}
After finding the configuration of the last $N_\eta$ elements, we update $\hat{\vect{d}}_\theta$ and $\hat{\vect{R}}_\theta$ and proceed to find the configuration of the elements $(\eta - 2)N_\eta+1$ to $(\eta-1)N_\eta$. To further clarify this sequential procedure, let $\tilde{\boldsymbol{\theta}}[n]$ be the $n$th RIS reflection coefficient obtained via this method for $n = iN_\eta+1,\ldots,N$. The problem of finding the configuration for the elements $(i-1)N_\eta+1$ to $iN_\eta$ is similar to the one in \eqref{eq:heuristic_SD} with $\hat{\vect{d}}_\theta$ and $\hat{\vect{R}}_\theta$ being updated as \eqref{eq:d_theta_update} and \eqref{eq:R_theta_update} as given in the top of the next page. 
\begin{figure*}[t]
    \begin{align}
        \label{eq:d_theta_update}&\hat{\vect{d}}_\theta = \Tilde{\vect{d}}_\theta \left[ (i-1)N_\eta +1: iN_\eta\right] - \Tilde{\vect{R}}_\theta \left[(i-1)N_\eta +1: iN_\eta,iN_\eta+1:N\right]\tilde{\boldsymbol{\theta}}\left[iN_\eta+1:N\right] \\
        &\hat{\vect{R}}_\theta =  \Tilde{\vect{R}}_\theta \left[(i-1)N_\eta +1: iN_\eta,(i-1)N_\eta +1: iN_\eta \right] \label{eq:R_theta_update}
    \end{align}
    \hrulefill
\end{figure*}
Algorithm \ref{Alg:Heuristic-SESD} summarizes the proposed heuristic SESD-based algorithm for optimizing the discrete RIS configuration. 

\begin{algorithm}[t!]
\caption{Heuristic SESD-based algorithm for solving \eqref{eq:generalized_SD}.}
\label{Alg:Heuristic-SESD}
\begin{algorithmic}[1]
\STATEx {\textbf{Inputs}: $\Tilde{\vect{d}}_\theta$, $\Tilde{\vect{R}}_\theta$, $N$, $\eta$. }
\STATE{Set $N_\eta = N/\eta$.}
\STATE{$\hat{\vect{d}}_\theta \leftarrow \tilde{\vect{d}}_\theta \left[ (\eta -1)N_\eta +1:N \right]$}
\STATE{$\hat{\vect{R}}_\theta \leftarrow \Tilde{\vect{R}}_\theta \left[(\eta -1)N_\eta +1:N, (\eta -1)N_\eta +1:N \right]$}
\STATE{Solve $\boldsymbol{\theta}^\prime = \argmin{\boldsymbol{\theta} \in \mathcal{D}^{N_\eta}}\,\, \|\hat{\vect{d}}_{\theta} - \hat{\vect{R}}_{\theta}\boldsymbol{\theta}\|^2$ via SESD}
\FOR{$i = \eta -1 : -1: 1$}
\STATE{$\tilde{\boldsymbol{\theta}}\left[iN_\eta+1 : (i+1)N_\eta \right] \leftarrow \boldsymbol{\theta}^\prime$}
\STATE{$\hat{\vect{d}}_\theta \leftarrow \Tilde{\vect{d}}_\theta \left[ (i-1)N_\eta +1: iN_\eta\right] - \Tilde{\vect{R}}_\theta \left[(i-1)N_\eta +1: iN_\eta,iN_\eta+1:N\right]\tilde{\boldsymbol{\theta}}\left[iN_\eta+1:N\right]$}
\STATE{$\hat{\vect{R}}_\theta \leftarrow \Tilde{\vect{R}}_\theta \left[(i-1)N_\eta +1: iN_\eta,(i-1)N_\eta +1: iN_\eta \right]$}
\STATE{Solve $\boldsymbol{\theta}^\prime = \argmin{\boldsymbol{\theta} \in \mathcal{D}^{N_\eta}}\,\, \|\hat{\vect{d}}_{\theta} - \hat{\vect{R}}_{\theta}\boldsymbol{\theta}\|^2$ via SESD }
\ENDFOR
\STATEx{ \textbf{Output:} $\tilde{\boldsymbol{\theta}}$.} 
 \end{algorithmic}
\end{algorithm} 

\begin{example}
Consider the following problem 
\begin{equation}
\label{eq:heuristic_example}
   \minimize{\vect{x} \in \mathcal{X}^{N}}\,\left\|\vect{d} - \vect{R}\vect{x}\right\|^2, 
\end{equation}where $\vect{d} \in \mathbb{R}^{N}$, $\vect{R} \in \mathbb{R}^{N \times N}$. The entries of $\vect{d}$ are distributed as $\mathcal{U}[0,20]$ and the upper-triangular entries of $\vect{R}$ are distributed as $\mathcal{U}[0,1]$. We assume that $N = 40$. The set of possible values for the entries of $\vect{x}$ is given by $\mathcal{X} = \{-1,1\}$. We consider $100$ random realizations of $\vect{d}$ and $\vect{R}$ and investigate four different scenarios for solving problem \eqref{eq:heuristic_example}: \begin{enumerate}
    \item  We solve \eqref{eq:heuristic_example} directly using the SESD algorithm.

 \item We employ the low-complexity heuristic SESD algorithm with $\eta = 4$ and $N_\eta = 10$.

\item We employ the low-complexity heuristic SESD algorithm with $\eta = 8$ and $N_\eta = 5$.

\item We employ the low-complexity heuristic SESD algorithm with $\eta = 10$ and $N_\eta = 4$.
\end{enumerate}

We compute the normalized mean square error (NMSE) of the value of the objective function in \eqref{eq:heuristic_example} as 

\begin{equation}
    \mathrm{NMSE} = \mathbb{E} \left\{ \frac{|q^\star - \bar{q}|^2}{|q^\star|^2}\right\},
\end{equation}where $q^\star$ denotes the objective function value when the problem is directly solved by the standard SESD algorithm and $\bar{q}$ is the objective function value when the heuristic SESD algorithm with either of the three above cases is applied for solving problem \eqref{eq:heuristic_example}. The expectation is over random realizations of $\vect{d}$ and $\vect{R}$.
Table~\ref{tab:NMSE} shows the corresponding NMSE values. We can see that NMSE is smaller for smaller values of $\eta$. This is because with a smaller $\eta$, a larger number of entries are simultaneously optimized in each sub-problem of the heuristic SESD method, making the obtained objective function value closer to its minimum value acquired by the standard SESD algorithm. Therefore, there is a trade-off between complexity and accuracy in the proposed heuristic SESD method. A smaller $\eta$ and consequently a larger $N_\eta$ can improve the solution accuracy at the cost of increased complexity. When $\eta = 1$, the heuristic SESD turns to the standard SESD where all the $N$ entries are jointly optimized. We can thus tune $\eta$ according to the required accuracy level and available computational resources. 

\begin{table}[!t]
\centering
\caption{NMSE of the objective function value of \eqref{eq:heuristic_example} when heuristic SESD is used.}
\begin{tabular}{|p{0.25\linewidth}| p{0.25\linewidth}|}
\hline 
\multicolumn{1}{|l|}{~Problem Setup}  &  ~~~~~~NMSE \\ \specialrule{.25em}{.07em}{.075em}
\multicolumn{1}{|l|}{~$\eta = 4,~N_\eta = 10$} &  ~~~~~~$0.0085$           \\ \hline
\multicolumn{1}{|l|}{~$\eta = 8,~N_\eta = 5$}    &   ~~~~~~$0.0181$            \\ \hline
\multicolumn{1}{|l|}{~$\eta = 10,~N_\eta = 4$}    &      ~~~~~~$0.0284$        \\ \hline
\end{tabular} \label{tab:NMSE}
\end{table}
\end{example}

\section{Complexity Analysis}
\label{sec:complexity}

In this section, we analyze the complexity of the WMMSE algorithm. We first examine the continuous RIS configuration scenario studied in Section~\ref{sec:proposed_WMMSE}. The complexity of Algorithm~\ref{Alg:WMMSE} mainly originates from computing $\vect{F}_k \vect{w}_i$ for $k,i = 1,\ldots,K$, the SESD algorithm for optimizing the precoding vectors, and the alternating optimization employed for computing the RIS reflection coefficients. The complexity of computing $\vect{F}_k \vect{w}_i$ is $O(MN)$, and as it has to be performed $K^2$ times, the total complexity is $O(K^2 MN)$. Then, the complexity of optimizing precoding vectors using the SESD algorithm is $O(K L^{2\gamma_1 M})$ for some $0 \leq \gamma_1 \leq 1$. Note that the value of $\gamma_1$ depends on the specific problem, e.g., the statistics of the channels, and there is no trivial way to compute the exact value of $\gamma_1$ \cite{jalden2005complexity}. As for the RIS configuration optimization, the main complexity corresponds to computing $\theta_n^* [\vect{A}]_{n,\hat{n}}$ for $n,\hat{n} = 1,\ldots,N$, which has the total complexity $O(I_1 N^2)$, where $I_1$ denoting the number of iterations in the alternating optimization. The total complexity of the entire algorithm is therefore $O\left(I_2(K^2MN + KL^{2\gamma_1 M} + I_1 N^2) \right)$, where $I_2$ is the number of iterations of the weighted sum MSE minimization algorithm.

In the case of discrete RIS configurations, we will have the same steps as in Algorithm~\ref{Alg:WMMSE}, except for step~12, where the alternating optimization method is replaced by the SESD algorithm for optimizing RIS reflection coefficients. The total complexity will be therefore $O\left(I_2(K^2MN + KL^{2\gamma_1 M} + 2^{\gamma_2 b N}) \right)$, where $O(2^{\gamma_2 b N})$ for some $0 \leq \gamma_2 \leq 1$ is the complexity of the SESD algorithm for solving the RIS configuration optimization problem. 

Finally, using our proposed low-complexity heuristic SESD algorithm for RIS configuration optimization, the complexity of optimizing the discrete RIS reflection coefficients is reduced to $O(\eta 2^{\gamma_2 b N/\eta})$ and the total complexity becomes $O\left(I_2(K^2MN + KL^{2\gamma_1 M} + \eta 2^{\gamma_2 b N/\eta}) \right)$. Table~\ref{tab:complexity} summarizes the complexity orders of the WMMSE algorithm for different kinds of RIS configurations.

\begin{table}[!t]
\centering
\caption{Complexity of the WMMSE algorithm in different RIS configuration scenarios.}
\begin{tabular}{|ll|}
\hline 
\multicolumn{1}{|l|}{RIS Configuration}  & ~~~~~~~~~~~~~~Complexity \\ \specialrule{.25em}{.07em}{.075em}
\multicolumn{1}{|l|}{Continuous} &    $O\left(I_2(K^2MN + KL^{2\gamma_1 M} + I_1 N^2) \right)$          \\ \hline
\multicolumn{1}{|l|}{Discrete (Optimal SESD)}    &   $O\left(I_2(K^2MN + KL^{2\gamma_1 M} + 2^{\gamma_2 b N}) \right)$               \\ \hline
\multicolumn{1}{|l|}{Discrete (Low-Complexity)}    &   $O\left(I_2(K^2MN + KL^{2\gamma_1 M} + \eta 2^{\gamma_2 b N/\eta}) \right)$               \\ \hline
\end{tabular} \label{tab:complexity}
\end{table}

\section{Numerical Results and Discussion}
\label{sec:Num_results}
In this section, we will evaluate the performance of the proposed algorithms in terms of convergence and sum rate. We will use two benchmarks for evaluating the performance of the SESD-based precoding and RIS configuration design:
\begin{enumerate}
\item \textbf{Coordinate descent}: In each iteration of the WMMSE algorithm, a coordinate descent approach is utilized for optimizing the discrete vector. Specifically, in this approach, entries of the vector of interest are iteratively optimized such that in each iteration, one of the entries is optimized using an exhaustive search over the discrete set while other entries are kept fixed. This approach has been used in \cite{Di2020,Wu2020b} for optimizing the discrete RIS configuration. 
 \item \textbf{Nearest point}: The optimal continuous solution is first found and then each entry of the optimized continuous vector is mapped to its nearest value in the discrete set (i.e., set of quantization alphabets or set of discrete reflection coefficients). In particular, in nearest point precoding,  we use the optimal infinite-resolution precoding vectors as given in \eqref{eq:infinite_precoding} in each iteration of the WMMSE algorithm and quantize the entries of the precoding vectors obtained after convergence. For nearest point RIS configuration, the alternating optimization approach presented in Section~\ref{sec:optimize_RIS} is employed in each iteration of the WMMSE algorithm and the obtained configuration after convergence is quantized. References \cite{Alexandropoulos2020,Peng2021reconfigurable,HZhang2022} have utilized this method for finding the discrete RIS configuration. 
\end{enumerate}
\subsection{Simulation Setup}
We consider an RIS-assisted MU-MIMO system where the following setup and parameters are used unless otherwise stated. The AAS at the BS is assumed to be a uniform linear array (ULA) equipped with $M = 8$ antennas and the RIS has a uniform planar array (UPA) structure with $N_{\mathrm{H}} = 8$ elements in the horizontal orientation and $N_{\mathrm{V}} = 8$ elements in the vertical orientation, and a total of $N = 64$ reflecting elements. The number of UEs is set as $K = 5$ and the number of quantization levels at the fronthaul is assumed to be $L = 4$. The channels from the BS to the RIS and from the RIS to the UEs are modeled by Rician fading. Specifically, we have
\begin{align}
 \label{eq:Rician_fading_channel}
    &\vect{H} = \sqrt{\rho_H}\left(\sqrt{\frac{\kappa_H}{\kappa_H + 1}} \vect{H}_{\mathrm{LOS}} + \sqrt{\frac{1}{\kappa_H + 1}} \vect{H}_{\mathrm{NLOS}} \right), \\
    &\vect{g}_k =\sqrt{\rho_{g,k}}\left(\sqrt{\frac{\kappa_g}{\kappa_g + 1}} \vect{g}_{\mathrm{LOS}} + \sqrt{\frac{1}{\kappa_g + 1}} \vect{g}_{\mathrm{NLOS}} \right)
\end{align}where $\kappa_H$ and $\kappa_g$ denote the Rician factors of the corresponding channels, set as $\kappa_H = \kappa_g = 5$ throughout the simulations, and the subscripts $\mathrm{LOS}$ and $\mathrm{NLOS}$ represent the line-of-sight (LOS) and non-LOS components of the channels. In particular,  $\vect{H}_{\mathrm{LOS}} = \vect{a}_{\mathrm{BS}}(\Omega) \vect{a}_{\mathrm{RIS}}^{\Ttran}(\varphi_{\mathrm{AoA}},\phi_{\mathrm{AoA}})$, where
\begin{align}
    &\vect{a}_{\mathrm{BS}}(\Omega) = [1,e^{j\psi_{\mathrm{B}}},\ldots,e^{j(M-1)\psi_{\mathrm{B}}}]^{\Ttran}, \\
    & \vect{a}_{\mathrm{RIS}}(\varphi_{\mathrm{AoA}},\phi_\mathrm{AoA}) = \big [1, e^{j\psi_{\mathrm{H}}},\ldots,e^{j\left((N_{\mathrm{H}} - 1)\psi_{\mathrm{H}} + (N_{\mathrm{V}} - 1) \psi_{\mathrm{V}}\right)} \big]^{\Ttran},
\end{align}
where $\vect{a}_{\mathrm{BS}}(\cdot)$ and $\vect{a}_{\mathrm{RIS}}(\cdot)$ are the array response vectors of the BS and the RIS, respectively, and the relative phase shifts are given by 
\begin{equation}
    \begin{aligned}
        \psi_{\mathrm{B}} &= \frac{2\pi}{\lambda}\delta_{\mathrm{B}} \sin(\Omega), \\
        \psi_{\mathrm{H}} &= \frac{2\pi}{\lambda}\delta_{\mathrm{H}}\sin(\varphi_{\mathrm{AoA}})\cos(\phi_{\mathrm{AoA}}), \\
        \psi_{\mathrm{V}} &= \frac{2\pi}{\lambda}\delta_{\mathrm{V}}\sin(\phi_{\mathrm{AoA}}), \notag
    \end{aligned}
\end{equation}
with $\lambda$ being the wavelength, $\delta_{\mathrm{B}}$, $\delta_{\mathrm{H}}$, and $\delta_{\mathrm{V}}$ denoting the spacing between the AAS antennas, RIS horizontal elements, and RIS vertical elements, and $\Omega$, $\varphi_{\mathrm{AoA}}$, and $\phi_{\mathrm{AoA}}$ indicating the angle of departure (AoD) from the BS, azimuth angle of arrival (AOA) to the RIS, and elevation AOA to the RIS, respectively. We set $\Omega = \pi/6$, $\varphi_{\mathrm{AoA}} = -\pi/3$, $\phi_{\mathrm{AoA}} = \pi/6$, and   $\delta_{\mathrm{B}} =\delta_{\mathrm{H}} = \delta_{\mathrm{V}} = \lambda/2$. For $\vect{H}_{\mathrm{NLOS}}$, we consider correlated Rayleigh fading with Gaussian distribution \cite{Demir2022Channel}. Furthermore, $\rho_{H}$ denotes the path-loss and is modeled as \cite{Emil2020RISvsDF}
\begin{equation}
     \rho_H = -37.5 - 22 \log_{10} \big(d_H/1~ \mathrm{m}\big) ~~[\mathrm{dB}],
\end{equation} with the carrier frequency of $3$\,GHz. $d_H$ is the distance between the BS and the RIS, which is set as $d_H=20\,$m. The channel $\vect{g}_k$ is modeled in a similar way with $\varphi_{\mathrm{AoD},k}$, $\phi_{\mathrm{AoD},k}$, and $d_{g,k}$ being the azimuth AoD from the RIS to UE\,$k$, elevation AoD from the RIS to  UE\,$k$, and the distance between UE\,$k$ and the RIS. We assume that the UEs are uniformly distributed around the RIS such that $d_{g,k} \sim \mathcal{U}[20\,\mathrm{m},40\,\mathrm{m}]$, $\varphi_{\mathrm{AoD},k} \sim \mathcal{U}[-\pi/3,\pi/3]$, and $\phi_{\mathrm{AoD},k} \sim \mathcal{U}[-\pi/6,\pi/6]$. The noise power spectral density is $-174$\,dBm/Hz and the bandwidth is $1$\,MHz. 

\begin{figure}
\centering
\subfloat[$N = 32$, $b = 2$.]
{\includegraphics[width=\columnwidth]{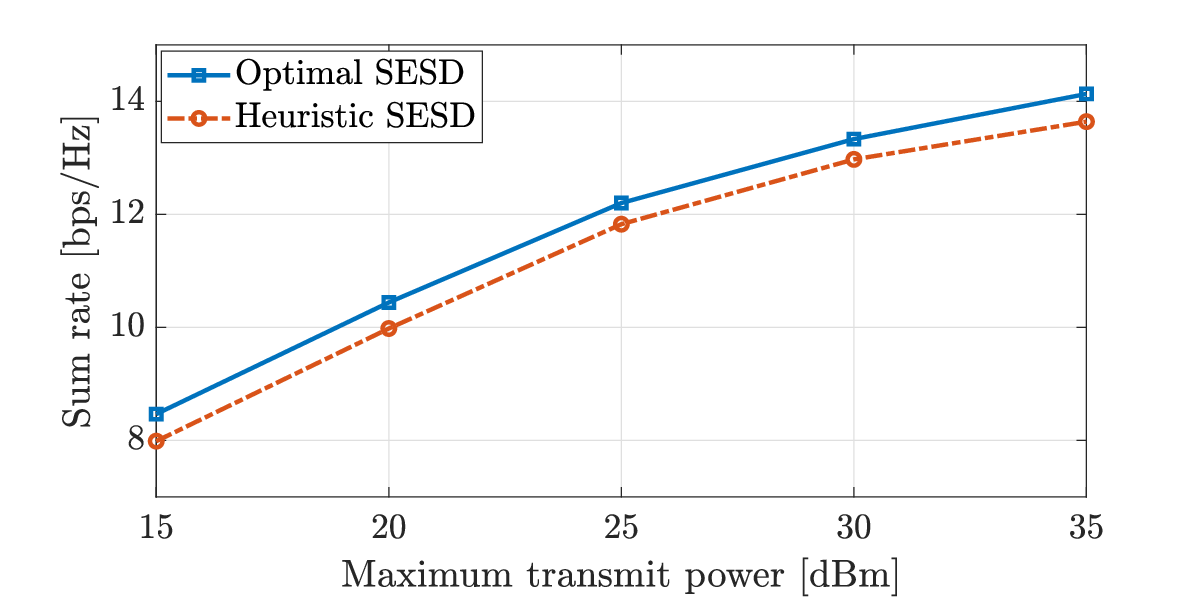}}\hfill
\centering
\subfloat[$N = 64$, $b = 1$.]
{\includegraphics[width=\columnwidth]{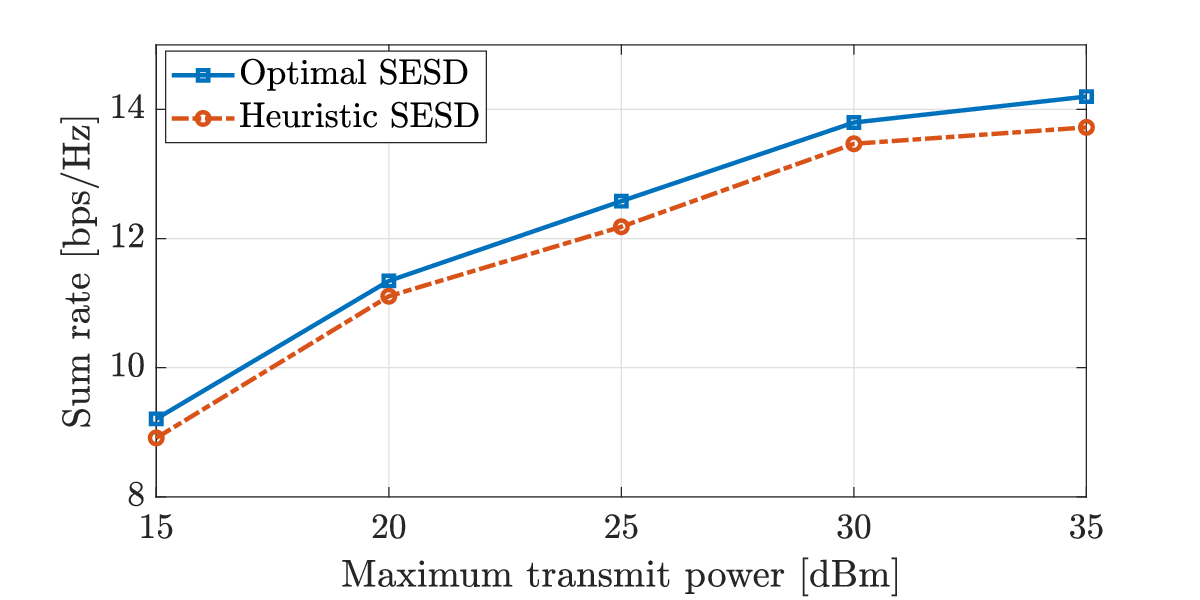}}
\caption{Sum rate vs. maximum transmit power for the case of discrete RIS configuration. The heuristic SESD algorithm performs relatively close to the optimal SESD algorithm while remarkably saving on complexity.}
\label{fig:SESDvsHeuristic}
\end{figure}

\begin{figure}[t]
    \centering
    \includegraphics[width = \columnwidth]{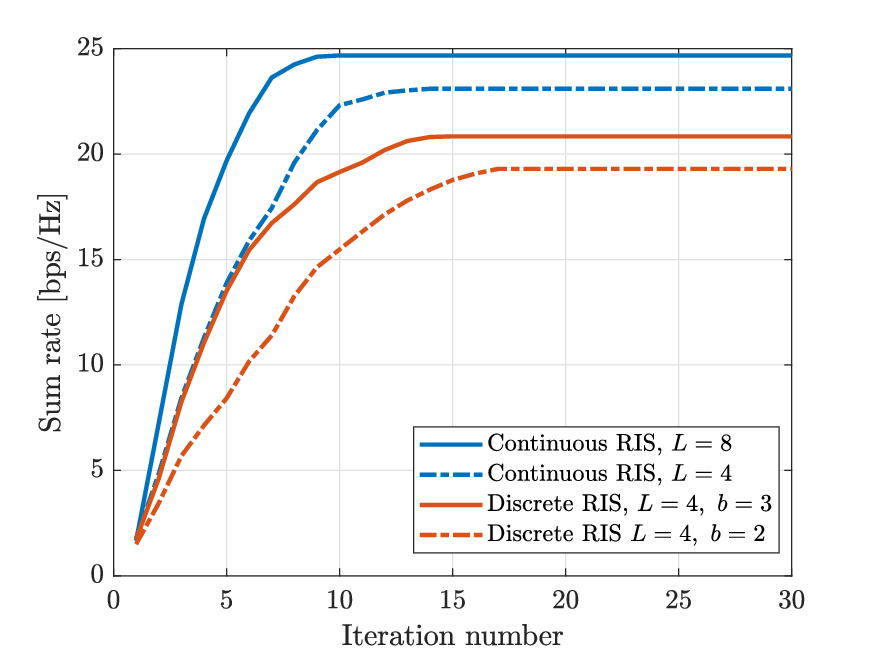}
    \caption{Sum rate evolution of the WMMSE algorithm for different number of quantization levels at the fronthaul and different bit resolutions at the RIS.} 
    \label{fig:Convergence}
\end{figure}

\subsection{Simulation Results}
We first evaluate the performance of the heuristic SESD-based algorithm proposed in Section~\ref{sec:heuristic_SESD} since this algorithm will be used for later simulations involving discrete RIS configuration. We consider two scenarios for the RIS setup, where in the first one, we assume that the RIS has $N = 32$ elements with $N_{\mathrm{H}} = 8,~ N_{\mathrm{V}} = 4$ and the bit resolution of the RIS is assumed to be $b = 2$. In the second scenario, we have $N = 64$ RIS elements in the form of a square UPA and the bit resolution is set as $b = 1$. We split the original RIS optimization problem into $\eta$ sub-problems such that $N_
\eta = 8$.  Therefore, we have $\eta = 4$ for $N =32$ and $\eta = 8$ for $N = 64$.
Fig.\,\ref{fig:SESDvsHeuristic} depicts the sum rate as a function of maximum transmit power for the optimal SESD design and heuristic SESD design. We can observe that the heuristic SESD design performs close to the optimal SESD design in both scenarios, while notably reducing the complexity.

Fig.~\ref{fig:Convergence} demonstrates the convergence behavior of the WMMSE approach presented in Algorithm~\ref{Alg:WMMSE}. The maximum power of downlink signals is set as $p = 30$\,dBm, and both continuous and discrete RIS cases are investigated. Specifically, for the continuous RIS case, two different number of quantization levels are considered: $L = 4$ and $L = 8$. For the discrete RIS scenario, the quantization level at the fronthaul is set fixed as $L = 4$ and two different bit resolutions at the RIS are considered: $b = 2$ and $b = 3$. The convergence of the continuous RIS case is faster than that of the discrete case because the obtained phase shifts in each iteration are closer to their global optimum value.
Furthermore, the figure shows that the convergence is achieved earlier with larger quantization level and bit resolution. This is due to the fact that the quantized precoding vectors and discrete RIS configuration are closer to their optimal unquantized ones with greater quantization level and bit resolution; thus, in each iteration, better precoding and RIS configuration solutions are obtained, thereby speeding up the convergence. For the very same reason, the achievable sum rate is higher with greater quantization level and bit resolution. Note that the faster convergence and larger sum rate are achieved at the expense of increased complexity and one should be mindful of this trade-off when choosing the appropriate number of available quantization levels and phase shifts.  

\begin{figure}[t]
    \centering
    \includegraphics[width = \columnwidth]{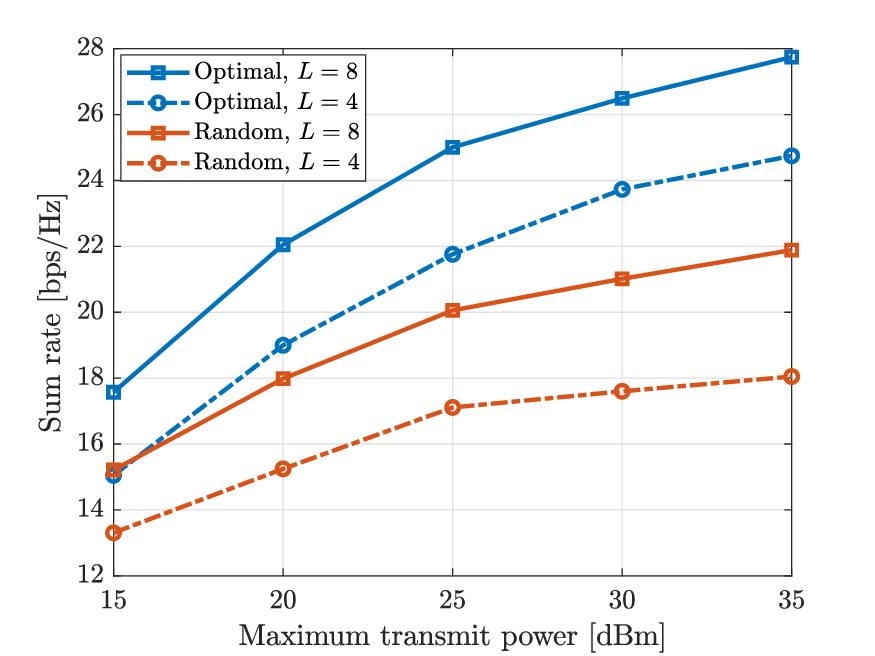}
    \caption{Sum rate vs. maximum downlink power when the RIS phase shifts are either optimized or randomized.  }
    \label{fig:OptimizedvsRandomRIS}
\end{figure}

Fig.~\ref{fig:OptimizedvsRandomRIS} investigates the benefits of using the RIS for assisting the communication between the BS and the UEs. The blue curves indicate the scenarios with optimized RIS phase shifts where the optimization is performed as discussed in Section~\ref{sec:optimize_RIS}, while the orange curves correspond to cases with random phase shifts which can represent scattering from a rough surface. In the latter case, the RIS phase shifts are once randomized, and the same values are then used throughout the simulations. Therefore, the problem reduces to optimizing the precoding vectors only. Two quantization levels for precoding is considered: $L = 4$ and $L = 8$. It is observed that using an RIS with optimized phase shifts can improve the system performance which highlights the gain that can be brought by RIS into the system.
\begin{figure}[t]
    \centering
    \includegraphics[width = \columnwidth]{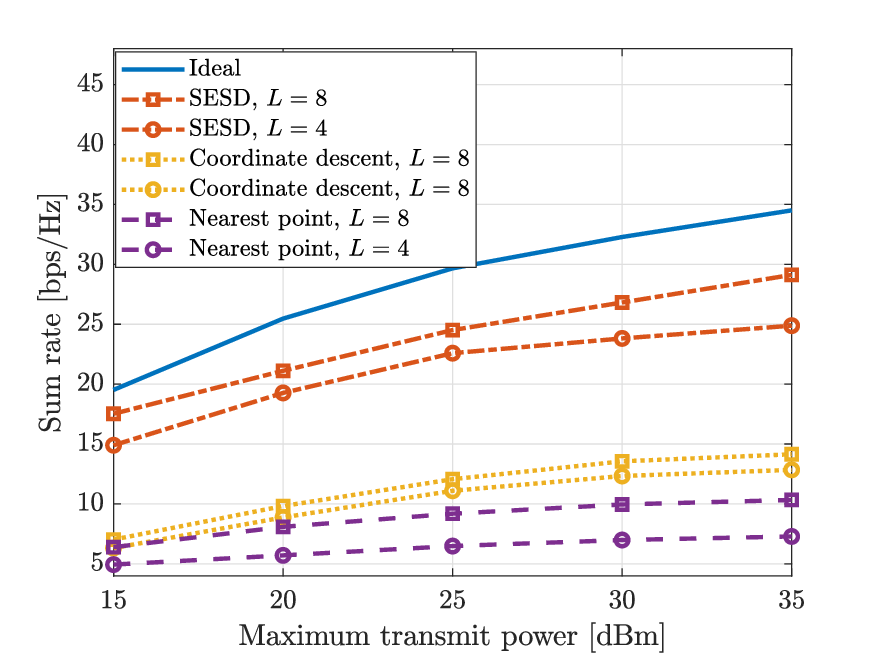}
   \caption{ Sum rate vs. maximum downlink power for the ideal case with infinite-resolution precoding and three different finite-resolution precoding optimization methods. }
    \label{fig:SESDvsNearest_Precoding}
\end{figure}

In Fig.~\ref{fig:SESDvsNearest_Precoding}, we evaluate the performance of the proposed SESD-based precoding by comparing it with nearest point precoding and coordinate descent precoding. In this simulation, continuous RIS configuration is considered and two quantization levels at the fronthaul are invesigated: $L = 4$ and $L = 8$. The ideal scenario with infinite-resolution precoding is also plotted which outperforms all the schemes with discrete precoding levels due to its higher interference suppression capability. We can see in Fig.~\ref{fig:SESDvsNearest_Precoding} that the sum rate of the SESD-based precoding is much superior to those of  nearest point precoding and coordinate descent precoding,  and the gap between the our SESD-based design and the benchmarks increases with increasing the transmit power. That is because the benchmark schemes fail to efficiently suppress the interference among UEs since they only obtain sub-optimal precoding vectors.
On the contrary, the SESD-based precoding which finds the optimal discrete precoding vectors enjoys performance improvement when a larger transmit power is used at the BS. The sum rate will saturate at high SNRs also with the proposed method, but at a much higher level that is not visible in this figure.

\begin{figure}[t]
    \centering
    \includegraphics[width = \columnwidth]{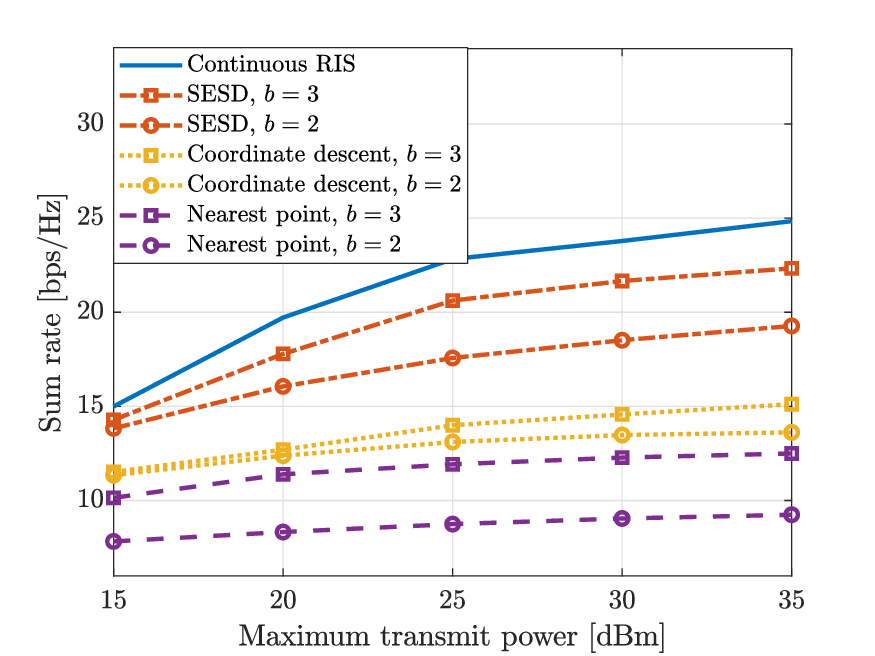}
    \caption{Sum rate vs. maximum transmit power for the continuous RIS case where the phase shift resolution of RIS is infinity and three different finite-resolution RIS optimization methods.} 
    \label{fig:SESDvsNearest_PhaseShift}
\end{figure}

Fig.~\ref{fig:SESDvsNearest_PhaseShift} presents the sum rate performance for the case of discrete RIS phase shifts, where the number of fronthaul quantization levels is assumed to be $L = 4$. Two bit resolutions at the RIS are considered: $b = 2$ and $b = 3$.
We compare the performance of the proposed SESD-based method with nearest point and coordinate descent approaches for finding the discrete RIS configuration. In both benchmarks, the precoding design is based on the optimal SESD algorithm.   
 The figure demonstrates the notable performance gain of the SESD-based configuration as compared to benchmarks. Furthermore, having a greater bit resolution leads to performance enhancement because with more available reflection coefficients, the optimized discrete configuration is closer to the rate-maximizing continuous configuration. 

An important observation can be made by comparing Fig.~\ref{fig:SESDvsNearest_Precoding} and Fig.~\ref{fig:SESDvsNearest_PhaseShift}. The gap between the SESD-based approach and the benchmarks is greater in Fig.~\ref{fig:SESDvsNearest_Precoding} than in Fig.~\ref{fig:SESDvsNearest_PhaseShift}. Although an RIS with continuous phase shifts is considered in Fig.~\ref{fig:SESDvsNearest_Precoding}, the performance of the benchmark schemes that employ sub-optimal precoding vectors is very poor. When an optimal precoding design is considered in Fig.~\ref{fig:SESDvsNearest_PhaseShift}, the benchmarks demonstrate a better performance, although the RIS phase shifts are still sub-optimally found in these schemes. This suggests that the  optimal precoder design is more important for improving performance as compared to the optimal RIS configuration design. The reason is that with the sub-optimal RIS configuration designs, only the phase shifts are affected as the reflection coefficients of the RIS are assumed to have unit modulus. On the other hand, a sub-optimal precoding design also affects the amplitude of the precoding entries, which has a large impact on the achievable sum rate.    

\subsection{Discussion}
In the previous subsection, we have seen the clear advantage of the SESD algorithm over the coordinate descent and nearest point methods in terms of achievable sum rate. However, this improved performance is attained at the expense of the higher complexity of the SESD algorithm. Although the heuristic SESD algorithm proposed in Section~\ref{sec:heuristic_SESD} can remarkably reduce the complexity compared to the standard SESD algorithm (as indicated in Table~\ref{tab:complexity}), this algorithm still has an exponential complexity, with respect to the number of RIS elements in each sub-problem..
\begin{table}
\centering
\caption{Average run time of heuristic SESD, coordinate descent, and nearest point methods for RIS configuration optimization. The numbers represent the run time in seconds.}
\begin{tabular}{|l||*{5}{c|}}\hline
\backslashbox{Method}{Bit Resolution}
&\makebox[4em]{$b = 2$}&\makebox[4em]{$b = 3$}
\\\specialrule{.25em}{.07em}{.075em}
Heuristic SESD & $30.98$ & $319.54$\\\hline
Coordinate Descent &$18.71$& $14.44$\\\hline
Nearest Point &$5.61$&$5.74$\\\hline
\end{tabular}\label{tab:RunTime}
\end{table}

To quantitatively compare the complexity of the heuristic SESD algorithm and the two considered benchmarks, we assess the average run time of solving the WMMSE problem for one realization of user locations and channels. It is assumed that the precoding optimization for all the three scenarios is performed via the standard SESD algorithm, where there are $L = 4$ quantization levels. For the RIS configuration optimization, we consider the three methods mentioned above. The transmit power is set as $p = 30$\,dBm.  Table~\ref{tab:RunTime} shows the run time of the WWMSE algorithm for $b = 2$ and $b = 3$ bit resolutions at the RIS. We can see that the run time of the proposed method is respectively $1.6$ and $5.5$ times that of the coordinate descent and nearest point methods for $b = 2$.  When the bit resolution is increased to $b = 3$, the proposed method is $22$ times and $55$ times slower than the coordinate descent and nearest point methods, respectively. The run time of the coordinate descent algorithm is reduced when the bit resolution is increased. This is due to the fact that with a finer phase shift grid at the RIS, the alternating optimization procedure involved in the WMMSE problem converges more quickly to a stationary point because the quantized phase shifts at each iteration are closer to their actual optimal values. The reward of the longer run time is the superior performance we achieve with the proposed method, as previously demonstrated in Fig.~\ref{fig:SESDvsNearest_PhaseShift}. As discussed in Section~\ref{sec:heuristic_SESD}, we can choose a larger value of $\eta$ to reduce the dimension of the RIS configuration optimization sub-problems, but this will also result in performance loss.  
Hence, optimality comes at the cost of increased complexity and we can reduce the complexity by sacrificing performance. 

It is worth noting that several methods have been proposed in the literature to reduce the complexity of SD algorithms with small performance losses. Interested readers can refer to \cite{Barbero2008Fixing,Ahn2009Schnorr,Ghasemmehdi2011} for more information.

\section{Concluding Remarks and Future Outlook}
\label{sec:conclusions}

The newly emerged RIS technology is regarded as a crucial player in the evolution of future wireless communications. While there has been considerable research in recent years on RIS-assisted communications, the focus has often been on simplified system models. This approach limits the direct application of theoretical findings to real-world scenarios.

We studied an RIS-assisted MU-MIMO downlink communication scenario in this paper. Since the precoding is computed in the cloud in future networks, the precoding vectors must be selected from a discrete set.
Furthermore, a practical RIS typically has a low phase shift resolution per element. 
We developed a novel framework for sum rate maximization considering precoding vectors and RIS configurations adhering to the above-mentioned limitations. We devised efficient solutions based on the SESD algorithm for both precoding and RIS configuration design and compared them against the well-known coordinate descent and nearest point designs. Numerical results demonstrated the effectiveness of the proposed designs and revealed the noticeable performance gains of the presented schemes over the commonly used benchmarks. The SESD complexity grows exponentially with respect to the problem size, making it unsuitable for practically large RIS. Hence, we developed a novel low-complexity SESD-based algorithm that divides the discrete RIS configuration optimization problem into a number of sub-problems, each solved via SESD. The effectiveness of the proposed low-complexity algorithm has also been corroborated numerically. This low-complexity approach can be utilized to solve any mixed-integer least-squares problem where the large problem size inhibits the use of conventional SD-based algorithms.  

Further exploration on how to reduce the complexity of SESD-based precoding and RIS configuration optimization with minimal performance loss is important. Methods such as lattice reduction and statistical pruning as well as fixed-complexity SD algorithms are worth a closer look in the context of discrete precoding and RIS optimization. There are also many other RIS-related scenarios where the discrete phase shifts might be optimized by following the methodology outlined in this paper.

\bibliographystyle{IEEEtran}
\bibliography{refs} 
\end{document}